\documentclass{jfm}

\usepackage{filecontents}

\usepackage{graphicx}
\usepackage{newtxtext}
\usepackage{newtxmath}
\usepackage{natbib}
\usepackage{xr-hyper}
\usepackage{hyperref}
\hypersetup{
    colorlinks = true,
    urlcolor   = blue,
    citecolor  = black,
}

\newcommand{\RomanNumeralCaps}[1]
\linenumbers

\usepackage[capitalise]{cleveref}
\crefformat{equation}{Eq.~#2#1#3}
\usepackage{flexisym}
\usepackage{booktabs} % Teja : for nice tables
\usepackage{makecell} % Teja : for multiline cells in a table
\usepackage{verbatim} % Teja : provides multiline comments
\usepackage{circledsteps} % Teja : To have a circled number.
\newcommand{\gv}[1]{\ensuremath{\mbox{\boldmath$ #1 $}}} % For vectors
\newcommand{\bv}[1]{\ensuremath{\boldsymbol{#1}}} % For tensors

\newcommand{\abs}[1]{\left\lvert#1\right\rvert}
\newcommand{\imag}[1]{\mathrm{Im} \left[ #1 \right]}
\newcommand{\real}[1]{\mathrm{Re} \left[ #1 \right]}
\newcommand{\RN}[1]{\textup{\uppercase\expandafter{\romannumeral#1}}}
\newcommand{\order}[1]{\mathcal O \left( #1 \right)}

\newcommand{\Er}{\ensuremath{Er}}

\newcommand{\dev}{\textprime}

\newcommand{\capsub}[1]{(\textit{#1})}
\newcommand{\todo}[1]{\textcolor{red}{TODO: #1}\PackageWarning{TODO:}{#1!}}
\newcommand{\ichang}[1]{\textcolor{blue}{#1}}

\newcommand{\appref}[1]{Supplementary Information {\cref{#1}}}
\renewcommand{\cite}[1]{\citep{#1}}

\makeatletter
\@input{xr_workaround.tex}
\makeatother

\title{A hyperelastic oscillatory Couette system}

\author{Tejaswin Parthasarathy\aff{1},
        Yashraj Bhosale\aff{1},
        \and
        Mattia Gazzola\aff{1} \corresp{\email{mgazzola@illinois.edu}}
 }

\affiliation{
\aff{1}Mechanical Sciences and Engineering, University of Illinois at Urbana-Champaign, Urbana, IL 61801, USA
}

\begin{document}
\maketitle
\begin{abstract}
%% Text of abstract
We report (semi-)analytical solutions of a problem involving a visco-hyperelastic solid material layer sandwiched
between two fluid layers, in turn confined by two long planar walls that undergo oscillatory motion. The resulting system dynamics are rationalized, based on fluid viscosity and solid elasticity, via wave and boundary-layer theory. This allows for physical interpretation of elasto-hydrodynamic coupling, potentially connecting to a broad set of biophysical phenomena and applications, from synovial joint mechanics to elastometry. Further, obtained solutions are demonstrated to be rigorous benchmarks for testing coupled incompressible fluid--hyperelastic solid and multiphase numerical solvers, towards which we highlight challenging parameter sets. Finally, we provide an interactive, online sandbox to build physical intuition, and open-source our code-base.
\end{abstract}


\section{Introduction}
\label{sec:intro}

\begin{figure}
\centering
\includegraphics[width=0.7\linewidth]{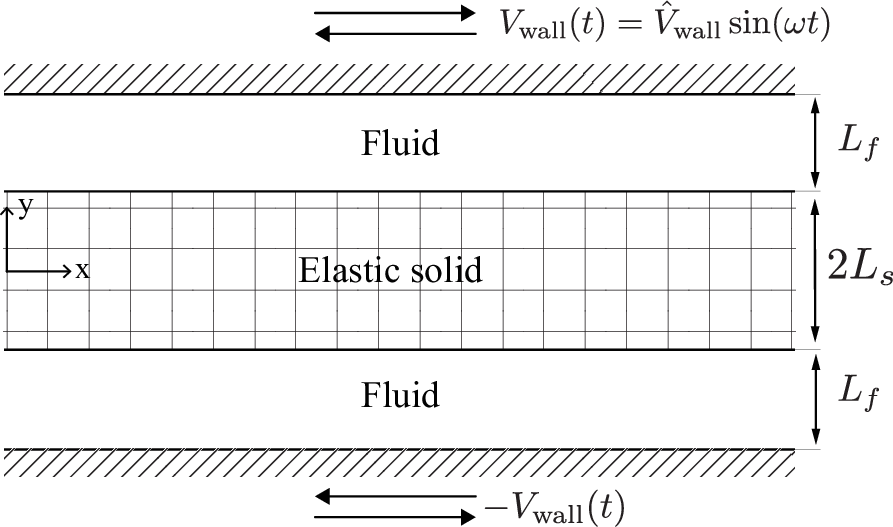}
\caption{\label{fig:setup}Schematic of the problem setup.}
\end{figure}

% In this paper, we investigate a benchmark case for testing coupled incompressible fluid--elastic solid
% interaction algorithms is described.
We report (semi-)analytical solutions of a minimal, yet representative problem involving a visco-hyperelastic solid material layer sandwiched between two fluid layers, in turn confined by two long planar walls that
undergo oscillatory motion~(\cref{fig:setup}). We are motivated by the ubiquity and relevance of coupled interactions between viscous fluid and visco-hyperelastic solids in engineering and biology~\cite{dowell2001modeling, grotberg2004biofluid,dong2011elastohydrodynamic,heil2011fluid,barthes2016motion}. Despite the numerous efforts to investigate this class of systems across modalities (theory, simulations, experiments) and applications, from vesicle transport~\cite{pozrikidis2003modeling,vlahovska2007dynamics},
%,barthes2016motion
pulmonary~\cite{grotberg2004biofluid,heil2008mechanics}, esophageal~\cite{kou2017simulation} or cardio-vascular systems~\cite{li2013continuum,bodnar2014fluid}, biolocomotion~\cite{argentina2007settling,gazzola2015gait,tytell2016role},
%geological flows~\cite{hewitt2015elastic,peng2020viscous},
microfluidics~\cite{wang2019theory,
%inamdar2020unsteady
christov2021soft},  % lubrication~\cite{hosoi2004peeling,skotheim2004soft},
drag reduction or energy harvesting~\cite{alben2002drag,alben2004flexibility,argentina2005fluid,bhosale2020bending}, there is a perhaps surprising paucity of rigorous, analytical benchmarks that capture, in a minimal setting, tightly coupled, interfacially-driven dynamics between hyperelastic solids and
shearing fluids. Such solutions are important to characterize relevant spatio-temporal scales, non-dimensional parameters and solution sensitivity, which are necessary for building intuition into practical flow--structure interaction problems.

Our setup, inspired by~\citet{sugiyama2011full}, caters to these requirements by coupling an incompressible Newtonian fluid to an incompressible density-mismatched visco-hyperelastic solid made of neo-Hookean and generalized Mooney--Rivlin materials, using a single, well-defined interface. By analyzing the flow field at this interface, we can understand the degree of dynamic coupling and mechanisms at play. This analysis is possible because in our setup the governing equations reduce to the simplest possible case of single dimension, while identically satisfying constraints of incompressibility. This results in decoupled algebraic equations which we solve to derive rigorous, analytical solutions. These solutions help isolate the spatio-temporal scales at play, and study parametric effects. % provide clarity on the governing non-dimensional parameters.

We begin by providing a detailed derivation of the flow solution, first in the case of a neo-Hookean solid, which in the limit of zero solid elasticity is shown to be consistent with classical multiphase Stokes--Couette flow solutions~\cite{landau1987theoretical,sim2006stratified,leclaire2014unsteady}. We then investigate the parametric impact of solid elasticity and fluid viscosity, and provide intuition for the observed results, using wave and boundary layer theory. These results are contextualized using applications in bio-engineering, from synovial joint mechanics to elastometry. During these parametric variations, we discover regimes marked by unusually high solid displacements, which we attribute to elastic, standing wave harmonics. We then carry forth our analysis from neo-Hookean solids to generalized Mooney--Rivlin solids, where closed-form solutions are no longer available. Here, to gain intuition into non-linear effects,
% we first conduct an asymptotic analysis  for small solid non-linearity by building upon our neo-Hookean solutions. We extend this intuition to large solid non-linearities, for which
we derive and analyze a semi-analytical solution.
\par
Our setup also serves as a useful benchmark for validating fluid--elastic structure interaction and multiphase simulation codes, towards which we highlight challenging parameter sets, and comparisons with direct numerical simulations~\cite{bhosale2021remeshed}. Finally, to further build intuition, we provide an interactive, online sandbox and open-source our code.

The work is organized as follows: the problem setup and governing equations are introduced in~\cref{sec:gov};
% governing equations are described in \cref{sec:gov};
 simplifications and analytical
solutions for neo-Hookean solids are discussed in \cref{sec:bmk2_derivation} with the
corresponding system behavior presented in~\cref{sec:nh_behavior}; ~\mbox{(semi-)}analytical solutions for Mooney--Rivlin materials and their interpretation are presented in \cref{sec:mr_solution};~concluding remarks are provided in \cref{sec:conc}.

\section{Problem setup and governing equations}\label{sec:gov}
A schematic of the setup is shown in \cref{fig:setup}, where we have a two-dimensional
visco-hyperelastic solid sandwiched between two layers of fluid, such that the
system is top-down symmetric. The thickness of the solid and each fluid layer is \( 2L_s\) and \( L_f\), respectively.
The setup is infinitely long and hence homogeneous in
the \(x\) direction. The fluid is bounded by two planar walls which
present a prescribed sinusoidal oscillatory motion \(V_{\textrm{wall}}(t) :=
  \hat{V}_{\textrm{wall}}\sin(\omega t) = \imag{\hat{V}_{\textrm{wall}} \exp(i
  \omega t) }\), where hatted quantities denote the Fourier coefficients
obtained upon a temporal Fourier transform, \( \omega\) is the angular frequency of oscillations, and \( T = 2\pi/\omega\) is the time period of oscillation. The bottom wall oscillates out of phase
with the top wall, with phase shift \(\pi\).
%
% We investigate the dynamics of the coupled solid--fluid system as a result of
% walls' oscillations. To this end, we first introduce the system governing equations, constitutive laws and boundary conditions that define the interactions between solids and fluids in our system.
%
% \section{Governing equations}\label{sec:gov}
%Here we present the governing equations, \subsection{Governing equations for solids and fluids}\label{sec:cauchy_gov}

\subsection{Governing equations}
We consider a two-dimensional domain \(\Sigma\) physically occupied by elastic solid
and a viscous fluid, with \(\Omega_{e}\) \& \(\partial\Omega_{e} \) representing the
support and boundaries of the elastic solid, respectively. % We then denote
% \(\overline{\Omega} = \overline{\Omega_{e}}\) as the solid material region, while the
The fluid region is represented by \(\Sigma - \overline{\Omega}\).

Linear and angular momentum balance in both the elastic solid and fluid phases on a
continuum scale leads to the Cauchy momentum equation
\begin{equation}
\label{eqn:cauchy1}
    \frac{\partial \gv{v}}{\partial t} + \bv{\nabla} \cdot
    \left( \gv{v} \bv{\otimes} \gv{v}\right) = -\frac{1}{\rho}\nabla p
    % + \gv{b}
    + \frac{1}{\rho} \bv{\nabla}
    \cdot \bv{\sigma}\dev ,~~~~\gv{x}\in\Sigma
\end{equation}
where \( t \in \mathbb{R}^+ \) represents time,
\(\gv{v} : \Sigma \times \mathbb{R}^+ \mapsto \mathbb{R}^2\) represents the velocity
field, \(\rho\) denotes material density, \(p : \Sigma \times \mathbb{R}^+
\mapsto \mathbb{R}\) represents the hydrostatic pressure field,
and \(\bv{\sigma}\dev : \Sigma \times \mathbb{R}^+
\mapsto \mathbb{R}^2 \otimes \mathbb{R}^2\) stands for the deviatoric
Cauchy stress tensor field. Throughout this work, the prime symbol \(\dev\) on a tensor $\bv{A}$
indicates its deviatoric, i.e.
\(\bv{A}\dev := \bv{A} - \dfrac{1}{2} {tr}(\bv{A}) \bv{I}\),
where \(\bv{I}\) stands for the identity tensor and
\({tr}(\cdot)\) represents the trace operator. All fields defined above are assumed to be
sufficiently smooth in time and space. Additionally, incompressibility of the
fluid and elastic domains results in the kinematic constraint on the velocity field
\begin{equation}
\label{eqn:incomp}
    \nabla \cdot \gv{v} \equiv 0,~~~~\boldsymbol{x}\in\Sigma.
\end{equation}
Interactions between the fluid and elastic solid phases take place via
interfacial boundary conditions, which correspond to continuity in velocities (no-slip)
and traction forces at the fluid--elastic solid interface
\begin{equation}
\label{eqn:elastic_bcs}
    \gv{v} = \gv{v}_f = \gv{v}_{e},~~~~
    \gv{n} \cdot \bv{\sigma}_f \cdot \gv{n} = \gv{n} \cdot \bv{\sigma}_{e}
    \cdot \gv{n},~~~~
    \gv{n} \cdot \bv{\sigma}_f \cdot \gv{t} = \gv{n} \cdot \bv{\sigma}_{e}
    \cdot \gv{t} ,~~~~\gv{x}\in \partial\Omega_{e}
\end{equation}
where $\gv{n}$ and $\gv{t}$ denote the unit outward (solid to fluid) normal
vector and the unit tangent vector at the interface \(\partial \Omega_{e}\), respectively.
Here, $\gv{v}_f$ and $\gv{v}_{e}$ correspond to the interfacial velocities in the fluid and
the elastic body, respectively, while $\bv{\sigma}_f = -p\bv{I} + \bv{\sigma}\dev_f$ and
$\bv{\sigma}_{e} = -p\bv{I} + \bv{\sigma}\dev_e$ correspond to the interfacial Cauchy stress tensor in the fluid
and the elastic body, respectively.

\subsection{Constitutive laws}\label{sec:closure}

In order to achieve closure of the above set of equations
(Eqs. \ref{eqn:cauchy1} to \ref{eqn:elastic_bcs}), %for the determination of the system dynamics,
we need to specify the form of internal material
stresses, i.e. the constitutive relations. In the following, we
discuss specific modeling choices for the deviatoric Cauchy stress tensor
\(\bv{\sigma}\dev\) of \cref{eqn:cauchy1}.%, for the different phases.

We assume the fluid to be Newtonian, isotropic and incompressible with density \(\rho_f\), dynamic
viscosity \(\mu_f\) and kinematic viscosity \(\nu_f = {\mu_f} / {\rho_f}\).
Accordingly, the Cauchy stress is defined as follows
\begin{equation}
\label{eqn:cons_fluid}
\bv{\sigma}_{f}{\dev} := 2 \mu_{f} {\bv{D}}\dev
\end{equation}
where \({\bv{D}}\dev\) is the strain rate tensor \(\frac{1}{2} \left ( \bv{\nabla}
\gv{v} + \bv{\nabla} \gv{v}^T \right)\).

Next, we assume that the elastic solid is isotropic, incompressible, has constant
density \(\rho_s\) and exhibits \emph{visco-elastic} behavior. Then the
deviatoric Cauchy stress can be defined as
\begin{equation}
\label{eqn:cons_solid}
\bv{\sigma}_{e}{\dev} := 2 \mu_{s} {\bv{D}}\dev + {\bv{\sigma}}\dev_{he},
\end{equation}
where \(\mu_{s}\) represents the dynamic
viscosity of the solid phase and
\({\bv{D}}\dev\) is the strain rate tensor. Similar to the fluid phase,
the kinematic viscosity of the solid is defined as \(\nu_s = {\mu_s} / {\rho_s}\).

The term \(\bv{\sigma}\dev_{he}\) corresponds to the
hyperelastic contribution to the deviatoric solid stress tensor. In this work,
\(\bv{\sigma}\dev_{he}\) is described through the generalized Mooney--Rivlin model~\cite{bower2009applied,sugiyama2011full},
appropriate to capture elastomeric and biological tissue responses
\begin{equation}
\label{eqn:solid_stress}
    \bv{\sigma}\dev_{he} :=
%\left( 2 \bv{F} \left[
% c_1 \frac{\partial \widetilde{\RN{1}}_{\bv{C}}}{\partial \bv{C}} +
% c_2 \frac{\partial \widetilde{\RN{2}}_{\bv{C}}}{\partial \bv{C}} +
% c_3 (\widetilde{\RN{1}}_{\bv{C}} - 2) \frac{\partial \widetilde{\RN{1}}_{\bv{C}}}{\partial \bv{C}}
% \right] \bv{F}^{T} \right)^\dev
\left(2 c_{1} \bv{B}+2 c_{2}({tr}(\bv{B}) \bv{B}-\bv{B} \cdot
\bv{B})+4 c_{3}({tr}(\bv{B})-2) \bv{B}\right){\dev},
\end{equation}
where \( \bv{B} \) is the left Cauchy--Green deformation tensor
\( \bv{B} := \bv{F}\bv{F}^{T}\), with \( \bv{F} \) defined as the deformation
gradient tensor \(\bv{F} := {\partial \gv{x}} / {\partial \gv{X}} \).
Here \(\gv{X}\) and \(\gv{x}\) correspond to the position of a material point
at rest and after deformation, respectively.

% In the infinitesimal deformations limit with \( c_3 = 0\), the entity \(2 \left( c_1 + c_2 \right)\) represents \(G\), the elastic shear modulus of the solid.
In the infinitesimal deformations limit, the entity \(2 \left( c_1 + c_2 \right)\) represents \(G\), the elastic shear modulus of the solid. Finally, setting \( c_2  = c_3 = 0\) and \( 2c_{1} = G \) in \cref{eqn:solid_stress},
results in the deviatoric Cauchy stress of a neo-Hookean material
\begin{equation}
\label{eqn:neo_hookean_stress}
\begin{aligned}
\bv{\sigma}\dev_{he} &= G \bv{B}{\dev}.
\end{aligned}
\end{equation}

Although the neo-Hookean model has been developed to capture % geometrically
non-linear stress-strain behaviours, it does so to a lesser degree of generality
relative to the generalized Mooney--Rivlin model~(\cref{eqn:solid_stress}).
Nonetheless, we consider here the neo-Hookean model as well, due to its simplicity,
popularity~\cite{bower2009applied,mihai2017characterize} and for comparison.

\section{Derivation of analytical solutions for neo-Hookean solids}\label{sec:bmk2_derivation}
% Next, we discuss in detail the derivation of the sharp interface (semi-)analytical solutions for
% this setup.
%
% With the governing equations established, we now look for their solutions in the
% context of our problem setup.
% \ichang{We first present a general solution technique, which follows the approach used
% in~\citet{sugiyama2011full}, but now accounts for viscosity and density mismatch
% in neo-Hookean and Mooney--Rivlin solids to establish a more meaningful
% biophysical connection.
% In the case of a neo-Hookean solid with linear governing equations, we first utilize this
% technique to obtain a closed-form analytical solution. Additionally, we present an alternative
% simpler, elegant analytical solution which we extend to the case of a generalized
% Mooney--Rivlin solid via asymptotic analysis. This
% asymptotic solution, valid only for small \( c_3 \) values, is finally used to contextualize a general semi-analytical solution,
% valid for arbitrary \(c_3\) values, which we derive from the technique described above.
% % For large \( c_3\) values we utilize the general solution technique presented above and
% % derive a .
% }
\subsection{Simplification of governing equations}
\label{sec:org0edba02}
We begin by noting that the problem is homogeneous in the \(x\)-direction and
hence we can omit the \(x\)-dependence of any quantity. % in the theoretical analysis that follows.
The problem then reduces to one-dimension, with gradients
only along the \(y\)-axis, and classical symmetry reductions to the governing
 Cauchy momentum equation can be adopted. First, the continuity equation (or equivalently the
incompressibility condition) of~\cref{eqn:incomp} simplifies to
\[  \frac{\partial v_y}{\partial y} \equiv 0. \]
A trivial solution to this equation is \( v_y(y, t) = c(t) \), where \( c(t) \) is an arbitrary
quantity. Because of the absence of motion of the wall in the y-direction,
\( c(t) \) is identically zero to match wall boundary conditions. Then, only the
displacement \(u(y,t)\), velocity \(v(y,t)\) and stresses \(\sigma_{xy} =
  \sigma(y,t)\) in the \(x\) direction need to be
considered. This simplifies the governing equation~\cref{eqn:cauchy1} to
read as
\begin{equation}
\label{eqn:simplified}
\begin{aligned}
    \partial_{t}u &= v, \\
    \rho \partial_{t}v &= \partial_{y}\sigma, \\
\end{aligned}
\end{equation}
where all quantities depend on \((y, t)\). Additionally, the stress can be
directly computed from \cref{eqn:cons_fluid,eqn:cons_solid,eqn:solid_stress}
as follows
\begin{equation}
\label{eqn:stress}
\sigma=\left\{\begin{array}{ll}
\mu_{f} \partial_{y} v & \text { for fluid }\left(L_{s} \leqslant |y| < L_{s}+L_{f}\right) \\
2c_{1} \partial_{y} u+4 c_{3}\left(\partial_{y} u\right)^{3} + \mu_{s} \partial_{y} v& \text { for solid }\left(0 \leqslant|y|<L_{s}\right)
\end{array}\right.
\end{equation}
where we remark that the coefficient \(c_2\) from~\cref{eqn:solid_stress} drops
out due to algebraic simplification and hence does not contribute to the dynamics.
We also note that \(c_3 = 0\) implies linear
stress responses with respect to the 1D displacement
$u$, while \( c_3 \neq 0 \) is responsible for (cubic) nonlinear behaviors~\cite{sugiyama2011full}.
The simplified~\cref{eqn:simplified,eqn:stress} indicate that accelerations (LHS)
result from viscous forces in the fluid, and a combination of viscous and elastic forces in the solid (RHS).

Further, velocities and stresses at the interfaces need to be continuous per \cref{eqn:elastic_bcs}, thus
\begin{equation}
\label{eqn:stress_continuity}
\begin{aligned}
v(\pm L_s^+, t) &= v(\pm L_s^-, t) \\
\mu_{f} \partial_{y} v(\pm L_s, t) &= 2c_{1} \partial_{y} u(\pm L_s, t) +4 c_{3}\left(\partial_{y}
u(\pm L_s, t)\right)^{3} + \mu_{s} \partial_{y} v (\pm L_s, t).
\end{aligned}
\end{equation}
Finally, we close the equations above by imposing no-slip
boundary conditions at the upper and lower walls at \(y = \pm (L_s + L_f)\)
\begin{equation}
\label{eqn:simplifed_bc}
v=\left\{\begin{array}{ll}
\hat{V}_{\textrm{wall}} \sin \omega t & \text { at } y=L_{f}+L_{s} \\
-\hat{V}_{\textrm{wall}} \sin \omega t & \text { at } y=-\left(L_{f}+L_{s}\right).
\end{array}\right.
\end{equation}
In the case of a neo-Hookean constitutive model (\(c_{3}\) = 0), the governing equations~\cref{eqn:simplified,eqn:stress,eqn:stress_continuity} reduce to a set of
linear equations since our setup involves purely shearing motions.
%apparent from the form of stress in~\cref{eqn:stress}.
We take two distinct, but
equivalent, solution approaches. The first one involves directly solving
the linear governing equations in the physical domain. The second one instead
solves the modal form of the governing equations obtained via a sine transform.
The first solution is elegant and compact, but only possible in the neo-Hookean case, while
the second solution is convoluted, but can handle arbitrary constitutive models.
We discuss both in the following.
% can be more general but more convoluted, is obtained by solving
% (in closed form) the linear modal~\cref{eqn:mom_fluid_modal,eqn:mom_solid_modal,eqn:stress_continuity_modal}
% for the expansion coefficients and reconstructing the physical velocity
% fields.
%
\subsection{Direct analytical solution}
\label{sec:phys_solution}
We first directly solve \cref{eqn:simplified,eqn:stress} in the fluid and solid domain. Given \( c_3 = 0\), we have
\begin{equation}
\label{eqn:reduced_stress}
\begin{array}{ll}
{\rho_f}\partial_{t}v_f = \mu_{f} \partial^2_{y} v_f & \text { for fluid }\left(L_{s}\leqslant y < L_{s}+L_{f}\right) \\
{\rho_s}\partial^2_{t}u_s = 2c_{1} \partial^2_{y} u_s + \mu_{s} \partial^2_{y}\partial_{t} u_s& \text { for solid }\left(0 \leqslant y <L_{s}\right).
\end{array}
\end{equation}
Considering the linearity of~\cref{eqn:reduced_stress}, symmetry of our setup, and the
sinusoidal form of wall velocity
\(V_{\textrm{wall}}(t) =  \imag{\hat{V}_{\textrm{wall}} \exp(i \omega t)}\),
one can expect similar sinusoidal forms in resulting displacements \( u_s(y, t) = \imag{\hat{u}(y) \exp (i \omega t) } \) and velocities \( v_f(y, t) = \imag{\hat{v}(y) \exp (i \omega t) }\).
Substituting these ansatzes in~\cref{eqn:reduced_stress} yields
\begin{equation}
\begin{array}{ll}
\left(\partial^2_{y} - \dfrac{i \omega}{\nu_f}\right) \hat{v}_f = 0 & \text { for fluid }\left(L_{s} \leqslant y < L_{s}+L_{f}\right) \\
\left(\partial^2_{y} + \dfrac{\omega^2}{i \omega \nu_s + \left({2c_1}/{\rho_s}\right) }\right) \hat{u}_s = 0 & \text { for solid }\left(0 \leqslant y <L_{s}\right)
\end{array}
\end{equation}
which are a pair of homogeneous Helmholtz equations with exact solutions
\begin{equation}
\begin{aligned}
\label{eqn:neo_exact}
\hat{v}_{f}(\tilde{y}) &= A \exp{\left( k_{f}\frac{\tilde{y}}{L_f} \right)} + B \exp{\left( -k_{f}\frac{\tilde{y}}{{L_f}} \right)} \quad &&\tilde{y} \in [0, L_f)  \\
\hat{u}_{s}(y) &= C \exp{\left( k_{s}\frac{y}{L_s} \right)} + D \exp{\left( -k_{s}\frac{y}{L_s} \right)} \quad && y \in [0, L_s),
\end{aligned}
\end{equation}
where
\begin{equation}
    \begin{aligned}
    \label{eqn:lambdas}
    k_{f} = {\sqrt{i}} \left( L_f^{-1} \left({\nu_f}/\omega\right)^{0.5} \right)^{-1} ,~~
    k_{s} = i \left[\left(\left(\omega L_s\right)^{-1}\left({2c_1}/{\rho_s}\right)^{0.5}\right)^2 + i \left(L^{-1}_s \left(\nu_s/\omega\right)^{0.5} \right)^2\right]^{-0.5}.
    \end{aligned}
\end{equation}
The coefficients \(A, B, C, D \) are directly determined given interface and boundary conditions~(Eqs. \ref{eqn:stress_continuity},~\ref{eqn:simplifed_bc}), and their expressions are reported in~\appref{app:neo_hookean_details}. Physically,~\cref{eqn:neo_exact} indicates a wave-like behavior, coupled with exponential decay, in both solid and fluid domains.
\subsection{Modal solutions using Fourier series}
\label{sec:fourier_series}
In the second approach, the solution strategy is to represent \(u(y,t), v(y,t)\) as a Fourier sine series
in the spatial coordinate \(y\), inject it into the governing equations \ref{eqn:simplified},~\ref{eqn:stress},
and match the interfacial conditions of~\cref{eqn:stress_continuity} and boundary conditions of~\cref{eqn:simplifed_bc} to obtain closed-form solutions. The choice of a sine series expansion is natural here given the Dirichlet velocity boundary conditions.
% \ichang{Additionally, it is possible to treat arbitrary constitutive models, besides the ones discussed here, with this strategy}.
Because of the piecewise definition of stresses in
\cref{eqn:stress} and interfacial condition in \cref{eqn:stress_continuity} (which indicates that velocities are \(\mathcal{C}^0\) continuous), convergence can be poor if a global Fourier series
(i.e. for both the solid and fluid domains together) is considered. Hence, we
utilize two piecewise Fourier series expansions for the solid \(v_s(y, t)\) and
fluid \(v_f(y, t)\) velocities, respectively, and explicitly impose \(\mathcal{C}^0\)
continuity in velocities and stresses.

We begin by noting that due to symmetry, \cref{eqn:simplified} can only admit solutions which are odd
functions of \(y\). Indeed, the equations of motion are invariant upon replacing \(u(y,t), v(y,t)\) with \(-u(-y,t), -v(-y,t)\). Then, one can expand \(u(y,t), v(y, t)\) using the Fourier sine series only in the upper half space \(y \geq 0\), as follows
\begin{equation}
\label{eqn:sine_series}
\begin{aligned}
    &v_{f}(\tilde{y}, t)=V_{I}(t)+\dfrac{\tilde{y}}{L_{f}}\left(V_{\textrm{wall}}(t)-V_{I}(t)\right)+\sum_{k=1}^{\infty} v_{f, k}(t) \sin \dfrac{\pi k \tilde{y}}{L_{f}} \\
    &u_{s}(y, t)=\dfrac{U_{I}(t) y}{L_{s}}+\sum_{k=1}^{\infty} u_{s, k}(t) \sin \dfrac{\pi k y}{L_{s}}
    % v_{s}(y, t)=\dfrac{V_{I}(t) y}{L_{s}}+\sum_{k=1}^{\infty} v_{s, k}(t) \sin \dfrac{\pi k y}{L_{s}},
\end{aligned}
\end{equation}
where \(U_{I}(t), V_{I}(t)\) is the displacement and velocity of the solid--fluid interface at \(y = L_s\), \(\tilde{y} = y - L_s\), and
\(v_{f,k}(t)\) and \(u_{s,k}(t)\) are the Fourier expansion coefficients of \( v_f \) and \(u_s\), respectively.
This expansion satisfies the incompressibility condition~(\cref{eqn:incomp}),
odd symmetry requirement about \( y = 0 \), interfacial velocity conditions (\cref{eqn:stress_continuity}), and boundary velocity conditions~(\cref{eqn:simplifed_bc}) imposed in the setup. Additional details regarding the expansion can be found in~\appref{app:piecewise}.
% , we motivate the choice of using additional piecewise linear functions to satisfy boundary conditions.
% (\( V_{I}(t)+\frac{\tilde{y}}{L_{f}}\left[V_{\textrm{wall}}(t)-V_{I}(t)\right] ; V_{I}(t) \frac{y}{L_{s}} \) in~\cref{eqn:sine_series})
% From \cref{eqn:sine_series}, one can deduce the solid displacement \(u_s\) as
% %
% \begin{equation}
% \label{eqn:solid_displacment}
% u_{s}(y, t)=\dfrac{U_{I}(t) y}{L_{s}}+\sum_{k=1}^{\infty} u_{s, k}(t) \sin \dfrac{\pi k y}{L_{s}}
% \end{equation}
We also note that the interface displacement \(U_{I}\) and modal displacement \(u_{s, k}\)  satisfy
\begin{equation}
\frac{\mathrm{d} U_{I}}{\mathrm{d} t}=V_{I}, \quad \frac{\mathrm{d} u_{s, k}}{\mathrm{d}
t}=v_{s, k}.
\end{equation}
Substituting the Fourier-series defined in \cref{eqn:sine_series} into the
governing~\cref{eqn:simplified} and utilizing
the stress relations of \cref{eqn:stress}, we rewrite
the equations with all terms moved to the LHS
\begin{equation}
    \begin{aligned} \label{eqn:mom_modal}
    \frac{\mathrm{d} V_{I}}{\mathrm{d} t}+\frac{\tilde{y}}{L_{f}}\left(\frac{\mathrm{d} V_{\textrm{wall}}}{\mathrm{d} t}-\frac{\mathrm{d} V_{I}}{\mathrm{d} t}\right)+\sum_{k=1}^{\infty}\left\{\frac{\mathrm{d} v_{f, k}}{\mathrm{d} t} + \nu_f\left(\frac{\pi k}{L_{f}}\right)^{2} v_{f, k}\right\} \sin \frac{\pi k \tilde{y}}{L_{f}} &=0 \\
    \frac{y}{L_{s}} \frac{\mathrm{d} V_{I}}{\mathrm{d} t} + \sum_{k=1}^{\infty}\left\{\frac{\mathrm{d}^{2} u_{s, k}}{\mathrm{d} t^{2}} + \nu_s\left(\frac{\pi k}{L_{s}}\right)^{2} \frac{\mathrm{d} u_{s, k}}{\mathrm{d} t} + \frac{2c_{1}}{\rho_s}\left(\frac{\pi k}{L_{s}}\right)^{2} u_{s, k} + \frac{\pi k}{\rho_{s} L_{s}} \sigma_{\mathrm{NL}, k}\right\} \sin \frac{\pi k y}{L_{s}} &=0
    \end{aligned}
\end{equation}
%
% \begin{equation}
% \begin{aligned}
% \label{eqn:mom_solid}
% &\frac{y}{L_{s}} \frac{\mathrm{d} V_{I}}{\mathrm{d} t} + \\ &\sum_{k=1}^{\infty}\left\{\frac{\mathrm{d}^{2} u_{s, k}}{\mathrm{d} t^{2}} + \nu_s\left(\frac{\pi k}{L_{s}}\right)^{2} \frac{\mathrm{d} u_{s, k}}{\mathrm{d} t} + \frac{2c_{1}}{\rho_s}\left(\frac{\pi k}{L_{s}}\right)^{2} u_{s, k} + \frac{\pi k}{\rho_{s} L_{s}} \sigma_{\mathrm{NL}, k}\right\} \sin \frac{\pi k y}{L_{s}}=0
% \end{aligned}
% \end{equation}
%
where $\nu_s = \mu_s / \rho_s$ and $\nu_f = \mu_f / \rho_f$ are the kinematic viscosities
of the solid and fluid phases, respectively. Here, \(\sigma_{\mathrm{NL}}\) denotes the
nonlinear contribution (i.e. the term containing \(c_3\)) in the solid stress \cref{eqn:stress} with respect to the
displacement, so that its expansion coefficients read
\begin{equation}
\label{eqn:nonlinear_stress}
\sigma_{\mathrm{NL}} := 4 c_{3}\left(\frac{\partial u_{s}}{\partial y}\right)^{3} =
\sum_{k=0}^{\infty} \sigma_{\mathrm{NL}, k} \cos \frac{\pi k y}{L_{s}}.
\end{equation}
We now project the governing equations in physical space (\cref{eqn:mom_modal}) onto
the Fourier modal bases, and then use Fourier identities
(\appref{app:fourier_identities}) to simplify the
obtained expressions
%The projected equations are shown below
\begin{equation}
\label{eqn:mom_fluid_modal}
\frac{2}{\pi k}\left\{\frac{\mathrm{d} V_{I}}{\mathrm{d} t}-(-1)^{k} \frac{\mathrm{d} V_{\mathrm{wall}}}{\mathrm{d} t}\right\}+\frac{\mathrm{d} v_{f, k}}{\mathrm{d} t}+\nu_{f}\left(\frac{\pi k}{L_{f}}\right)^{2} v_{f, k}=0
\end{equation}
\begin{equation}
\label{eqn:mom_solid_modal}
-\frac{2(-1)^{k}}{\pi k} \frac{\mathrm{d} V_{I}}{\mathrm{d} t}+\frac{\mathrm{d}^{2} u_{s, k}}{\mathrm{d} t^{2}} + \nu_s\left(\frac{\pi k}{L_{s}}\right)^{2} \frac{\mathrm{d} u_{s, k}}{\mathrm{d} t} + \frac{2c_{1}}{\rho_s}\left(\frac{\pi k}{L_{s}}\right)^{2} u_{s, k}+\frac{\pi k}{\rho_s L_{s}} \sigma_{\mathrm{NL}, k}=0
\end{equation}
with \(k = 1, \dots, \infty\). In modal space, the continuity condition of shear stresses at
the interface, upon substituting~\cref{eqn:sine_series} into~\cref{eqn:stress_continuity} and using the Fourier identities of~\appref{app:fourier_identities}, reads
\begin{equation}
\label{eqn:stress_continuity_modal}
\begin{aligned}
&\frac{\mu_{f}\left(V_{\textrm{wall}}-V_{I}\right)}{L_{f}} - \frac{2c_{1}U_{I}}{L_{s}} - \sigma_{\mathrm{NL}, 0} \\
&- \frac{\mu_{s}V_{I}}{L_{s}}
\sum_{k=1}^{\infty}\left[\frac{\mu_{f} \pi k v_{f, k}}{L_{f}}-(-1)^{k}\left\{\frac{2c_{1}
\pi k u_{s, k}}{L_{s}} +  \frac{\mu_{s}\pi k}{L_s}\frac{\mathrm{d} u_{s,k}}{\mathrm{d} t}+
\sigma_{\mathrm{NL}, k} \right\}\right] = 0.
\end{aligned}
\end{equation}

\Cref{eqn:mom_fluid_modal,eqn:mom_solid_modal,eqn:stress_continuity_modal}
directly relate the modal expansion coefficients \( v_{f, k} \) in the fluid, and
\( u_{s,k} \) in the solid,
via the interfacial quantities \(U_{I} \), \( V_{I}\) as a function of the physical setup parameters. We now need to solve~\cref{eqn:mom_fluid_modal} to \cref{eqn:stress_continuity_modal} for the modal quantities \( v_{f, k}, u_{s,k} \)
and interfacial quantities \(U_{I}, V_{I}\). To do so, we truncate the number of modes in the above infinite Fourier series to \(k = K - 1\). This
leads to a truncation error, which we minimize by taking \(K\) to be large. Here,
\(K\) is fixed to 1024 unless otherwise indicated.
\par
%
% \section{Analytical solutions for a neo-Hookean solid}
% \label{sec:neo_hookean_solution}
% \subsection{Modal solutions}
% \label{sec:modal_solution}
% In the case of a neo-Hookean constitutive model, \(c_{3}\) in \cref{eqn:stress}
% is identically 0. In this case, the governing equations
% reduce to a set of linear equations since our setup involves purely shearing
% motions, apparent from the form of stress in~\cref{eqn:stress}.
% The modal \cref{eqn:mom_fluid_modal,eqn:mom_solid_modal,eqn:stress_continuity_modal} can then be solved
% analytically for the expansion coefficients.
We now specialize the above solutions for the neo-Hookean case with \( c_3 = 0\). First,
similar to~\cref{sec:phys_solution}, we expect sinusoidal forms for the temporal quantities
\begin{equation}
    \label{eq:temporal_form}
    %\begin{aligned}
    V_{I}(t)    = \imag{\hat{V}_{I} \exp (i \omega t)},~~
    v_{f, k}(t) = \imag{\hat{v}_{f, k} \exp (i \omega t) },~~
    u_{s, k}(t) = \imag{\hat{u}_{s, k} \exp (i \omega t) }.
    %\end{aligned}
\end{equation}
% %
% with the immediate implication that
% %
% \begin{equation}
%     \label{eq:temporally_transformed}
%     \begin{aligned}
%         U_{I}(t)    &= \imag{\frac{\hat{V}_{I}}{i \omega} \exp (i \omega t)} \\
%         v_{s, k}(t) = \frac{\mathrm{d} u_{s, k}}{\mathrm{d} t} &= \imag{i \omega \hat{u}_{s, k} \exp (i \omega t) } \\
%         \frac{\mathrm{d}^{2} u_{s, k}}{\mathrm{d} t^{2}} &= \imag{-\omega^2 \hat{u}_{s, k}
%         \exp (i \omega t) }.
%     \end{aligned}
% \end{equation}
% %
%
Substitution of the temporal transformed quantities from~\cref{eq:temporal_form} in the momentum
ODEs (Eqs.~\ref{eqn:mom_fluid_modal}, \ref{eqn:mom_solid_modal}) leads to algebraic equations that
can be solved. Upon algebraic manipulation and taking into account the modal stress
balance (\cref{eqn:stress_continuity_modal}), we obtain
\begin{equation}
    \label{eq:linear_v}
    \begin{aligned}
        \hat{u}_{s, k}=-\frac{i(-1)^{k} \hat{V}_{I} \beta_{k}}{\pi \omega k}
        ,~~
        \hat{v}_{f, k}&=\frac{\left\{(-1)^{k} \hat{V}_{\mathrm{wall}}-\hat{V}_{I}\right\} \alpha_{k}}{\pi k}
     \end{aligned}
\end{equation}
where \( \hat{V}_{I}, \alpha_k, \beta_k\) are coefficients whose expressions are tedious and hence deferred to~\appref{app:neo_hookean_modal_details}. The expressions of~\cref{eq:linear_v} can then be directly used in~\cref{eqn:sine_series} to analytically evaluate solid displacements, fluid velocities and solid velocities. This provides the final modal solution for the case of a neo-Hookean solid.
\par
Our solution approaches are equivalent and generate the same results (\appref{app:comparison}). Having discussed both these approaches, we now identify key non-dimensional quantities that physically characterize the system, validate our solutions against known special cases and direct numerical simulations, analyze parametric behavior and investigate implications on the system response.
\section{Analysis of system behavior for neo-Hookean solids}\label{sec:nh_behavior}
\subsection{Key driving parameters}
\begin{table}
  \begin{center}
% \centering
    \begin{tabular}{lll}
    \toprule
    Symbol & Definition & Physical interpretation\\
    \midrule
    \(L\) & \(2(L_{s} + L_{f})\) & Length scale\\
    \(L_{f}/L_{s}\) &  & Length ratio \\
    \(\dot{\gamma}\) & \({2\hat{V}_{\mathrm{wall}}}/{\omega L}\) & Non-dimensional shear rate \\
    \(Re \) & \({\dot{\gamma} \omega L_f^2}/{\nu_f}\) & Reynolds number \\
    \(\Er \) & \( {\mu_f \dot{\gamma} \omega}/{2c_{1}}\) & Ericksen number \\
    \(\rho \) & \( {\rho_s}/{\rho_f}\) & Density ratio \\
    \(\nu \) & \( {\nu_s}/{\nu_f}\) & Viscosity ratio \\
    \(\delta_{f} \) & \( {L_f}^{-1}\left(\nu_f / \omega\right)^{0.5} = \exp\left( i \pi / 4\right) k_f^{-1} = \left(\dfrac{\dot{\gamma}}{\Rey}\right)^{0.5} \) & \makecell[cl]{Non-dimensional fluid \\ Stokes layer thickness}\\
    \(\delta_{s} \) & \( {L_s}^{-1}\left(\nu_s / \omega\right)^{0.5} = \sqrt{\imag{-k^{-2}_s}} = \left( L_f / L_s \right) \left(\dfrac{\nu\dot{\gamma}}{\Rey}\right)^{0.5} \) & \makecell[cl]{Non-dimensional solid \\ Stokes layer thickness} \\
    \(\lambda \) & \( {\left(\omega L_s\right)}^{-1}\left(2c_1 / \rho_s \right)^{0.5} = \sqrt{\real{-k^{-2}_s}} = \left( L_f / L_s \right) \left( \dfrac{\dot{\gamma}^2}{\rho \Rey \Er }\right)^{0.5}\) & \makecell[cl]{Non-dimensional elastic \\wavelength} \\
    \bottomrule
    \end{tabular}
    \caption{Characteristic non-dimensional parameters}
    \label{tab:params}
  \end{center}
\end{table}
The proposed system can be fully characterized through a set of non-dimensional parameters, deduced from our solutions above, which are listed in~\cref{tab:params}. Here the Reynolds number \Rey~captures the importance of inertial effects in the fluid phase using the ratio of inertial to viscous forces. Higher \Rey~indicates an inertia-dominated response from the fluid. The Ericksen number \( \Er \) captures the importance of elasticity in the solid phase using the ratio of viscous to elastic forces. Lower \( \Er \) indicates an elasticity-dominated response from the solid. The fluid Stokes-layer thickness \( \delta_f \) captures the boundary layer length scale associated with the exponential decay of wall velocity, relative to the fluid layer thickness. Low values of \( \delta_f \) indicate significant decay of wall velocity before reaching the interface. The solid Stokes-layer thickness \( \delta_s\) has a similar interpretation, but for the solid layer. The elastic wavelength \( \lambda \) captures the length scale associated with elastic shear waves progressing from the interface into the solid bulk, relative to the solid layer thickness. Low values of \( \lambda \) indicate high number of elastic-waves within the solid phase. The relevance of these length scales \( \delta_f, \delta_s, \lambda\) will become apparent as we discuss the system response in~\cref{sec:limit}.

In this work, we fix the geometry \( L_{s} = L_{f} = L / 4\), and unless stated otherwise, we assume  non-dimensional shear rate
\( \dot{\gamma} = \pi^{-1}\)
% \( \dot{\gamma} = 1 \)
and density ratio \( \rho = 1\). We remark that these assumptions do not affect the generality of our results. Indeed, the effects of \( \dot{\gamma}, \rho \) can be reproduced through fluid viscosity \(\nu\) and solid elasticity \(c_1\), via the non-dimensional parameters of~\cref{tab:params}.
\par
Having defined the key non-dimensional parameters, we can now investigate their impact on the behavior of the system.
\subsection{Limit cases}\label{sec:limit}
In order to develop a physical intuition for the system response, we first selectively remove the effects of solid viscosity (\( \nu_s \to 0\)) and elasticity (\(  c_1 \to 0 \)) and analyze our solution. In these limit cases we recover classical analytical results.

\subsubsection{Purely elastic solid case (\(\nu_s = 0\))}
\label{sec:pure_elastic}
\begin{figure}%[htbp]
\centering
\includegraphics[width=\textwidth]{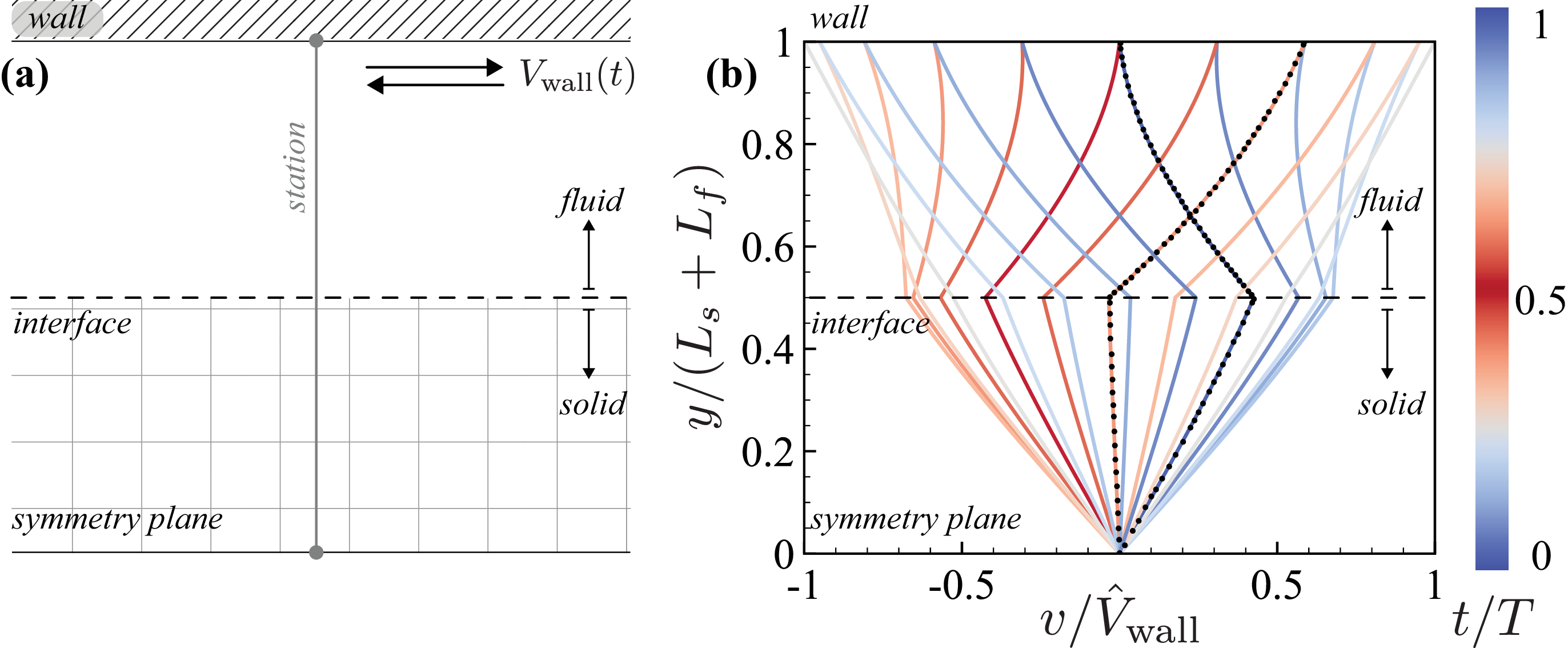}
\caption{\label{fig:linear_velocity} Pure elastic solid limit. Non-dimensional velocity profiles in \(y\) for a pure elastic solid with a neo-Hookean constitutive model. The system response is shown only in the upper half plane. The system is characterized by \(Re = 0.25 \), \(\Er = 1/\left(5\pi\right) \), \(\nu = 0\), \(\delta_f = 1.12\), \(\delta_s = 0\), \(\lambda = 2.52\). Additional details can be found in~\appref{app:fig_details}.
The same set of parameters are analyzed in~\citet{sugiyama2011full}, whose profiles (only provided at two time instants) are overlaid as black scatter points on our curves. Colors represent \(t / T\).}
\end{figure}

In the limit of \(\nu_s = 0\), the solid is purely elastic and we recover the solution of~\citet{sugiyama2011full} (up to minor typographical errors in that work). Here we consider the setup shown in~\cref{fig:linear_velocity} with parameters taken from~\citet{sugiyama2011full}, to enable comparison with their results. This system is characterized by \(Re = 0.25, \Er = 1/\left(5\pi\right), \nu = 0, \delta_f = 1.12, \delta_s = 0, \lambda = 2.52\).

We begin our comparison by highlighting typical non-dimensional velocity profiles obtained from our solutions. We only showcase profiles in the upper half-plane, shown in \cref{fig:linear_velocity}a, due to symmetry in our setup. We plot the profiles corresponding to the marked line station in \cref{fig:linear_velocity}a, at different time instants (or equivalently phases) within one oscillation cycle. These profiles are presented in \cref{fig:linear_velocity}b, with colors indicating time instants. For reference, the wall is located at \( y / (L_s + L_f) = 1\) and the symmetry plane is located at \( y / (L_s + L_f) = 0\). The interface is located at \( y / (L_s + L_f) = 0.5\), below which we have the elastic solid zone and above which we have the fluid zone. In this plot we also overlay the velocity profiles (black points) predicted by~\citet{sugiyama2010full}. While they provide profile data for only two time instants, we find favorable agreement with our velocity profiles, at both these times. From these velocity profiles, we see that the solid velocity exhibits a phase lag (indicated by the colors) relative to the fluid velocity, and less pronounced magnitudes. The fluid's maximal velocity magnitude always occurs at the wall (\( \abs{v} = \hat{V}_{\textrm{wall}} \)) while the solid velocity magnitudes always reach a minimum at the symmetry plane (\( \abs{v} = 0 \)). Finally, the slopes of the velocity profiles \( \partial_{y}v\) are discontinuous at the interface, to satisfy continuity in stresses (\cref{eqn:stress_continuity}).

We can gain an intuition for these profiles by considering force balance in the fluid and solid phases separately. That is, at any point in space-time, the sum of all real and apparent (i.e. inertial acceleration) forces must add up to zero. In the viscous fluid, we have inertial and viscous contributions, as seen from \( -\partial_t v + \nu \partial_y^2 v  = 0 \). This balance equation indicates that viscous forces operate by acting on the \emph{curvature} \( \partial_y^2 v \) of the velocity profile. Thus, both high viscosity \( \nu \) (low \( \Rey \)) and high velocity profile curvature \( \partial_y^2 v \) contribute to increasing viscous forces, which then exactly balance out accelerations. Typically, these viscous forces (and velocity profile curvatures) are concentrated within a boundary layer close to the wall (seen from the structure of the solution in \cref{eqn:neo_exact}), characterized by the non-dimensional Stokes layer thickness \( \delta_f \). Within this boundary layer, viscous forces cause the flow velocity to rapidly decay before eventually reaching the interface.

From the moving interface (no-slip), the solid phase displacement propagates into the bulk, mediated by elastic forces. From~\cref{eqn:simplified,eqn:stress}, the elastic contribution to solid force balance is \( -\partial_t v + 2c_1 \partial_y^2 u  = 0 \). This indicates that elastic forces operate by acting on the \emph{curvature} \( \partial_y^2 u \propto \omega^{-1} \partial_y^2 v\) of the solid velocity profile \( v\). So both high elastic shear modulus \( 2c_1 \) (low \(\Er\)) and high velocity profile curvature \( \partial_y^2 v \) contribute to increasing elastic forces. These elastic forces propagate as waves within the solid (\( \nu_s = 0\), so \( \lambda_2 \) from~\cref{eqn:lambdas} is purely imaginary, leading to sinusoids in \cref{eqn:neo_exact}), characterized by the non-dimensional elastic wavelength \( \lambda \). This implies that a wave profile can be expected for velocities (and curvatures) within the solid, which then always adjusts to zero at the symmetry plane in a fashion similar to \emph{nodes} in stationary waves. Additionally, for a viscoelastic solid (\( \nu_s \neq 0 \)), we have viscous effects that set up a boundary layer close to the interface and symmetry planes, similar to the fluid phase. The extent of this region is characterized by the non-dimensional solid Stokes layer thickness \( \delta_s\).

Overall, across both fluid and solid phases, we can rationalize the observed velocity profiles by considering \Rey, \Er~and the curvature length scales \( \delta_f, \delta_s, \lambda \). Referring back to~\cref{fig:linear_velocity}, since \( \Rey = 0.25 \sim \order{1}\), we expect inertial and viscous forces to be approximately equally important in the fluid. Additionally, \(\delta_f = 1.12 > 1\) indicates that the boundary layer occupies most of the fluid zone. This leads to moderate velocity curvatures throughout the fluid phase, as seen in \cref{fig:linear_velocity}b. This, in turn, drives the solid phase characterized by no viscosity and low \( \Er \), indicating stiff/strong elastic behavior. As a consequence of low \( \Er \), the wavelength \( \lambda \propto \Er^{-0.5} \) is large (\(\lambda = 2.52 > 1\)). We then expect to see only the nascent part of a wave, which is almost linear, as indeed observed in \cref{fig:linear_velocity}b.
\subsubsection{No elastic solid (\(c_1 = 0\)): single phase and multi-phase Stokes--Couette flows}
Our solution recovers classical results in the limit of \(c_1 = 0\), which indicates absence of elastic forces in the solid phase. Thus, only viscous forces operate in the solid, effectively rendering it a Newtonian fluid. If \( c_1 = 0 \), \(\mu_s = \mu_f = \mu \), and \( \rho_s = \rho_f = \rho \), then the entire domain is occupied by a single fluid, and we recover the Stokes--Couette flow solution~\cite{landau1987theoretical} valid throughout the domain. If instead \( c_1 = 0 \) but now \(\mu_s \neq \mu_f\) or \( \rho_s \neq \rho_f\), then the domain is occupied by two different fluids, and we recover the multi-phase Stokes--Couette flow for two immiscible liquids, which has established piecewise analytical
solutions~\cite{sim2006stratified,leclaire2014unsteady}. Upon comparison with single- and double-phase reference Stokes--Couette flow solutions, our analytical formulations (\cref{sec:phys_solution,sec:fourier_series}) are found in excellent agreement~(\appref{app:limit_cases}).
% The reference solutions, their comparison with our theory, and the physical interpretation of the results are reported in~\cref{app:limit_cases}.}.
%
\subsection{Neo-Hookean solid: Verification against numerical simulations}
\label{sec:nh_validation}
\begin{figure}%[htpb!]
\centering
\includegraphics[width=\textwidth]{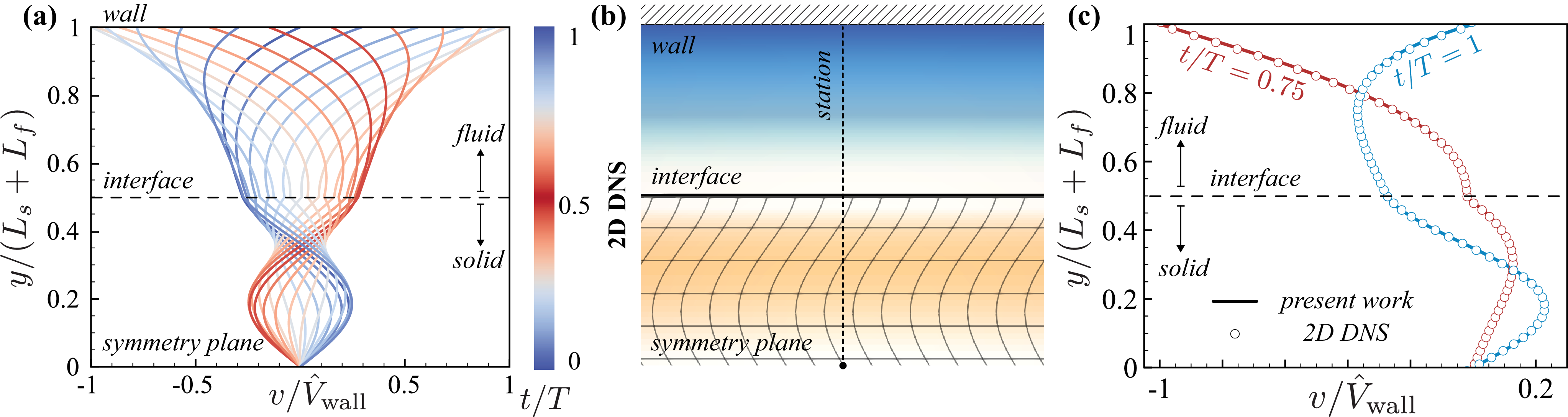}
\caption{
    \label{fig:neo_hookean_simulations}Comparison against simulations.
    \capsub{a} Non-dimensional analytical velocity profiles in \(y\), for an visco-elastic solid with a neo-Hookean constitutive model. The system response is shown only in the upper half plane.  This system is characterized by \(Re = 2, \Er = 1, \nu = 0.1, \delta_f = 0.4, \delta_s = 0.126, \lambda = 0.225\).
    Additional details can be found in~\appref{app:fig_details}.
    % Other parameters used are \(L_s = L_f = L / 4 = 0.2 \), \(\rho_f = \rho_s = 1\), \(\mu_f = 0.02\), \(\mu_s = 0.1\mu_f\), \(c_{1} = 0.01\), \(c_{3}=0\), \(\hat{V}_{\mathrm{wall}} = 0.4\), \(\omega = \pi\).
    Colors indicate \( t / T\). \capsub{b}
    %2D DNS setup: we feed in the parameter set above with the physical setup of \cref{fig:setup} in to simulations and run it until it reaches dynamical steady state, till $t/T = 10$,
    We solve the above problem through 2D DNS~\cite{bhosale2021remeshed} and run our simulations until the system reaches dynamical steady state ($t/T = 10$),
    and then sample quantities within the last cycle. Corresponding simulation parameters are found in~\appref{app:fig_details}. In the image, we mark the \(x\)-velocity field (orange/blue represent positive/negative velocity) and  deformation contours within the solid, with the interface marked (black, thick solid) for visual clarity. \capsub{c} Upon plotting the velocity profiles at the highlighted station (black, dashed) in the center of the domain, we see good agreement with our analytical results across all times. For the sake of clarity, we only show profiles at two different time instances. Here numerical results are plotted with scatter points whereas analytical results are plotted with a solid line.
}
\end{figure}

We now move on from analyses of limit cases and consider the more general scenario of visco-hyperelastic Neo-Hookean solids. Before studying the system behavior in a range of conditions (\cref{sec:neo_hookean_dynamics}), we validate our solutions against direct numerical simulations employing a recent two-dimensional remeshed-vortex method framework~\cite{bhosale2021remeshed}. In ~\cref{fig:neo_hookean_simulations}a, we consider a system characterized by \(Re = 2, \Er = 1, \nu = 0.1, \delta_f = 0.4, \delta_s = 0.126, \lambda = 0.225\). In these conditions \( \Rey, \Er \sim \order{1}\) so that we expect elastic, viscous and inertial
forces to be equally important in the solid, marking a departure from the above limit cases. Additionally, since \( \lambda < 1\), we expect the emergence of wave-like profiles inside the solid. As illustrated in~\cref{fig:neo_hookean_simulations}a, our analytical solutions within the solid indeed exhibit a standing wave-like behavior, constrained by boundary layer adjustments (with characteristic high curvatures) both near the interface and the symmetry plane. Further, these results are confirmed by our direct simulations as illustrated in~\cref{fig:neo_hookean_simulations}c, where we report the numerically obtained velocity profiles along the line-station of~\cref{fig:neo_hookean_simulations}b, overlaid on the theoretical predictions. As can be seen, profiles compare favorably at multiple temporal instants, validating the accuracy of both our theory and numerical solver.
%
% delta_f : 0.3989422804014327
% delta_s : 0.126156626101008
% lambda 0.22507907903927657
\subsection{Range of soft, elastomeric interface dynamics}
\label{sec:neo_hookean_dynamics}
Having validated our analytical solutions across different scenarios, we next investigate the dynamic response of the system for variations in the two most important parameters: elasticity (\( \Er \)) and viscosity (\(\Rey\)). Here we span the set of \( \Er = [0.1, 1, 10] \), which includes the range of soft cellular tissue found in the human body~\cite{wu2018comparison,guimaraes2020stiffness} and \(\Rey = [0.1, 0.5, 1, 2, 10]\), which indicates small to moderate inertial effects. Our choices capture typical values found in oscillatory micro-fluidic assays and applications involving biological and soft elastomeric materials that operate in conjunction with fluid interfaces~\cite{di2009inertial,velve2010microfluidic,duncombe2015microfluidics}.
\par
Of further relevance in the context of our minimal setup, we also find in-situ studies of bacterial deposition on coated, elastic surfaces in pulsatile flows~\cite{bakker2003bacterial}. In these cases, preferential adhesion based on elastic stiffness has been reported~\cite{song2015effects}, offering avenues to modulate bio-film formation and preventing bio-fouling. Our model may offer insights for the manipulation of oscillatory flow-stresses through soft elastic coatings~\cite{gad2002compliant}. Another potential application connects to the mechanics and wear of loaded human synovial joints~\cite{dowson1986micro,sun2003expression,nalim2004oscillating,sun2010mechanical}, where wall-driven, cyclic (synovial) fluid shear stresses act on soft articular cartilages. Finally, our model may be of use % in engineering for the in-situ analysis of elasto-hydrodynamic effects such as lubrication~\cite{dowson1986micro,skotheim2004soft,shinkarenko2009validity} as well as
in non-destructive testing of solid rheological properties, similar to Couette visco- and elasto-meters~\cite{carr1976physical}.
\begin{figure}%[htpb!]
\centering
\includegraphics[width=\textwidth]{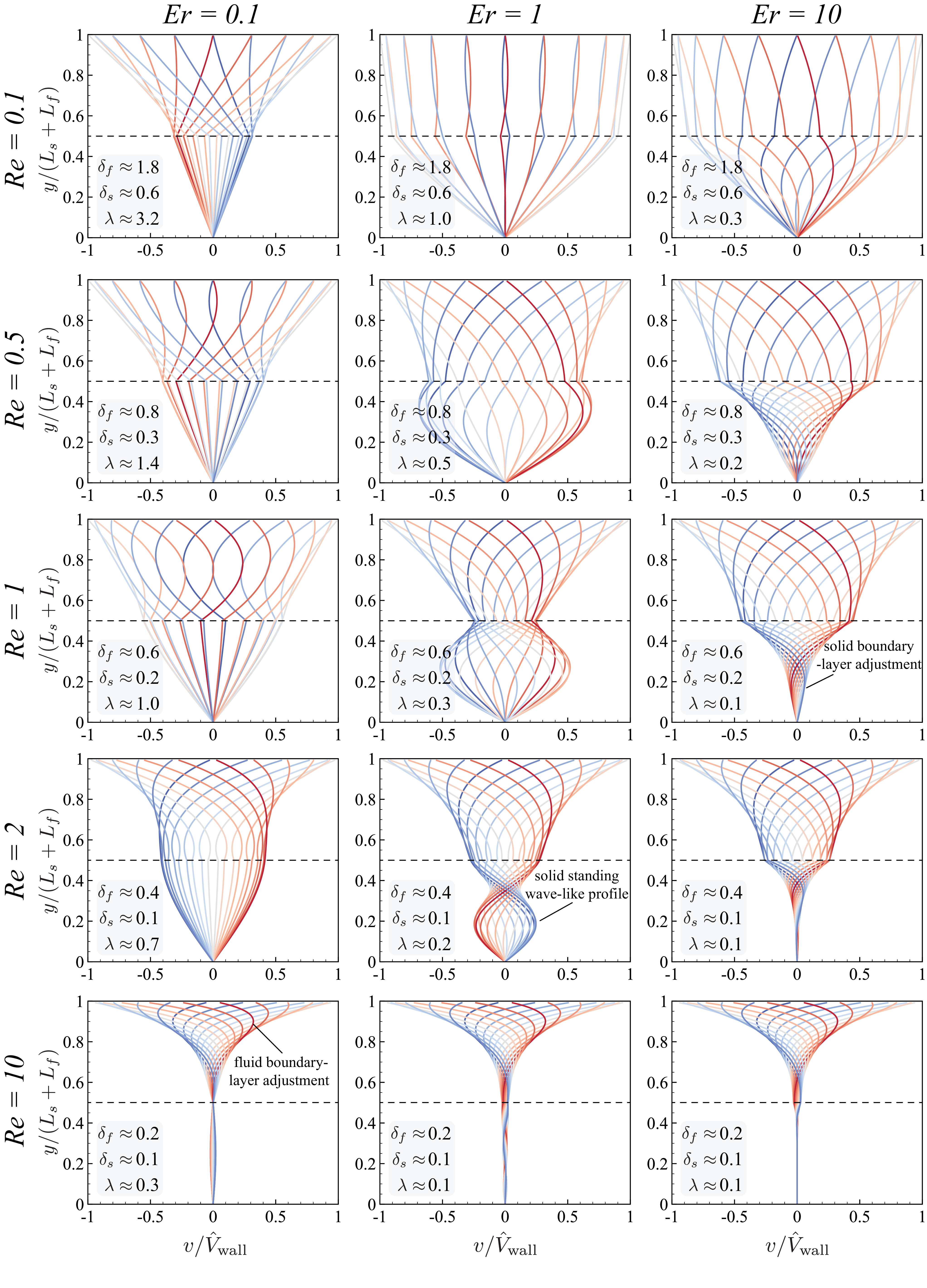}
\caption{\label{fig:neo_hookean_sweep}Dynamics with parametric variation. Non-dimensional velocity profiles in \(y\) for parametric changes in \( \Rey = [0.1, 0.5, 1, 2, 10] \; \times \; \Er = [0.1, 1, 10]\) with \( \nu = 0.1, \rho = 1\).
The system response is shown only for the upper half plane, where we also mark the length scales \( \delta_f, \delta_s, \lambda \). The black, dashed lines indicate the solid--fluid interface. Colors represent \(t / T\), with the same colorbar as \cref{fig:linear_velocity}.}
\end{figure}
\par
Within this context, we focus on the system response~(velocity profiles) first at low \( \Rey = 0.1 \) (top row of~\cref{fig:neo_hookean_sweep}), then at (relatively) high \( \Rey = 10\) (bottom row) and finally at intermediate \( \Rey \) (middle rows). For each \( \Rey \), we consider the impact of \( \Er \), ranging from stiff (low \(\Er\), left column) to soft solids (high \( \Er\), right column). Within each \( \Rey \) regime, we discuss the fluid velocity profiles first, followed by the solid velocity profiles. Velocity profiles are rationalized using the length scales \( \delta_f = \left(\dot{\gamma} \Rey\right)^{-0.5}, \delta_s = \left(\nu\dot{\gamma}/\Rey\right)^{0.5} \) and \( \lambda = \dot{\gamma}\left(\Rey \Er\right)^{-0.5}\), which are marked alongside each case study. % in~\cref{fig:neo_hookean_sweep}.
\subsubsection{Low \( \Rey \)}
First, at low \( \Rey = 0.1 \), we expect viscous forces to be important in the bulk flow. Indeed, the boundary layer in the fluid zone is characterized by \( \delta_f \approx 1.8 > 1\), so that its thickness spans the entire flow domain. This thick boundary layer results in two prominent effects. First, it indicates that fluid velocities have minimal curvature, which we confirm from the top row of~\cref{fig:neo_hookean_sweep}. Additionally, it effectively transmits the viscous stresses induced by the wall to the interface, which then initiates motion in the bulk solid. At this low \( \Rey \), the thickness \( \delta_s \approx 0.6\) of the solid boundary layer spans the bulk of the solid domain itself. In addition, unlike the fluid phase, we now also have elastic contributions, which we investigate by spanning \( \Er \). At low \( \Er = 0.1\), the solid has a large elastic wavelength \( \lambda \approx 3.2 > 1\). Hence, similar to~\cref{sec:pure_elastic}, we expect the velocity profiles to have no wave-like behavior, and thus less curvature. This is confirmed by the left column of~\cref{fig:neo_hookean_sweep}, where we see approximately linear solid velocity profiles, intuitively justified by the fact that to balance out acceleration forces, the elasticity modulus \( c_1 \) has to be large when curvatures are minimal (see~\cref{sec:pure_elastic}).

If we then increase \Er, going from stiffer (left column) to softer solids (right column), we expect both elastic and viscous forces to equally contribute to the dynamics. This is accompanied by decreasing values of \(\lambda\) which indicate that more wavelengths can now fit in the solid layer thickness. Then, similar to~\cref{sec:nh_validation}, we expect the appearance of standing wave-like profiles, with prominent boundary-layer adjustments close to interface and symmetry planes. These considerations are confirmed in \cref{fig:neo_hookean_sweep}, where we see that solid velocity profiles exhibit increasing curvatures as we move from left to right in the top row.
\subsubsection{Higher \( \Rey \)}
Next, for \( \Rey = 10\), we see prominent boundary layer adjustments in the fluid close to the wall. This is due to the characteristically low boundary layer thickness \( \delta_f \approx 0.2 < 1\), which implies that fluid velocity curvatures can be high only within this compact region. Indeed, beyond this boundary layer, the fluid velocity decays rapidly before reaching the interface, leading to the profiles shown in the bottom row of~\cref{fig:neo_hookean_sweep}. As a result of this decay, the flow cannot effectively transmit viscous stresses to the interface and hence the solid barely deforms. This flow decay is dependent only on \( \Rey \), and so we expect similar small solid deformation amplitudes even if we vary the solid elasticity. We confirm this intuition by increasing \( \Er \) (left to right), noticing small solid velocity amplitudes. Hence, in this \( \Rey \) regime, the fluid evolves almost independently (``weak coupling") from the details of the solid.
% elasticity, a phenomenon which we loosely refer to as ``weakly coupled" flow--structure interaction.
In contrast, the low \( \Rey \) regime seen earlier is ``strongly coupled". % had both fluid and solid evolving conjointly leading to a system.
Finally, we note that increasing \( \Er \), i.e.~decreasing \( \lambda \), leads to wavy profiles (although of small magnitude) inside the solid.
\subsubsection{Intermediate \( \Rey \)}
For intermediate \( \Rey\), the system showcases a rich variety of behaviors, which we highlight by investigating parameters around \( \Rey = 1\). Firstly, in these cases the fluid's boundary layer has moderate thickness \( \delta_f \sim \order{1}\) and hence we expect moderate velocity profile curvatures over \( \delta_f\). By decreasing \( \delta_f\) (e.g. by increasing \( \Rey \)), we expect the flow curvature to  increase. We confirm this in~\cref{fig:neo_hookean_sweep}, as we move from \( \Rey = 0.5 \) (top) to \( \Rey = 2\) (bottom). An increase in \( \Rey \) also increases the solid velocity curvatures, by decreasing both the solid wavelength \( \lambda \) and solid boundary layer thickness \( \delta_s \sim \order{0.1} \). The effect of decreasing \( \lambda \) is prominently displayed as we move from \( \Rey = 0.5 \to 2 \) for a fixed \( \Er = 1\) (central column). Further, at high \( \Er\), we expect viscous forces to dominate over elastic forces, thus rendering the solid medium more fluid-like. Indeed, for \( \Er = 10\), the solid velocity profiles showcase a boundary-layer adjustment similar to the one encountered in fluids. Hence, as we span \( \Er \) from 0.1 to 10 at intermediate \( \Rey = 0.5-2\), the effects of viscosity and elasticity compete in the solid leading to rich dynamics. As a consequence, in this regime, solid velocity profiles are especially sensitive to changes in \( \Er\). Such sensitivity provides a potential mechanism to manipulate and control interfacial stress magnitudes in the previously mentioned applications. Finally, because of its dynamic variety and sensitivity, this intermediate parameter regime is identified as numerically challenging, and therefore we propose the parameter set \( \Rey = \{0.5, 1, 2\} \) and \( \Er = \{0.1, 1, 10\}\) for benchmarking flow--structure interaction solvers, as illustrated in~\cref{fig:neo_hookean_simulations}.

\subsection{Solid phase resonance}\label{sec:gains}
We conclude this section by investigating the conditions under which resonant solid deformations occur. These may serve well for applications such as elastometry, where high amplitude peaks provide unique footprints to characterize materials.
% \ichang{Additionally, applications involving sensitive solids need to avoid near-resonant conditions for preventing material failure, which further motivates us to characterize regimes with large solid deformations}\todo{TEJA : I feel the above blue parts goes well here, but your comment seems to indicate otherwise Mattia. So I don't know where do you think it should fit best}.
%
\par
We begin by defining the gain function \( |G| \geq 0 \) as the ratio of solid to wall amplitude which, from~\cref{eqn:neo_exact}, takes the closed-form expression
\begin{equation}
|G| = \Biggl\lvert  \frac{i \omega C}{\hat{V}_{\textrm{wall}}} \Biggr \rvert = \frac{2}{\biggl\lvert  \left(e^{k_f-k_s} - e^{-(k_f-k_s)}\right) \left( 1 - \alpha \right) - \left(e^{k_f + k_s} - e^{-(k_f + k_s)}\right) \left( 1 + \alpha \right) \biggr \rvert}
\end{equation}
where \( k_f \) and \( k_s \) are the fluid and solid wave contributions~(\cref{eqn:lambdas}), and \( \alpha = \frac{L_f}{L_s} \frac{k_s}{k_f} \left( \rho \nu - i \frac{\dot{\gamma}}{\Er} \right) \) captures the degree of fluid--solid coupling.
\par
In the limit case of a purely elastic solid (\( \nu = 0\)) the denominator of \( G \) is always \( > 0 \), due to the non-zero contributions from the fluid phase (\( k_f \neq 0 \)). The immediate implication is that unbounded resonance \( |G| \to \infty \), is not possible in our setup because the fluid always dampens out high amplitudes in the solid phase. Thus, interstitial fluids, beside providing lubrication as in synovial joints, may also prevent excessive deformations and subsequent failure of the soft, articular cartilage.
\begin{figure}%[htpb!]
\centering
\includegraphics[width=\textwidth]{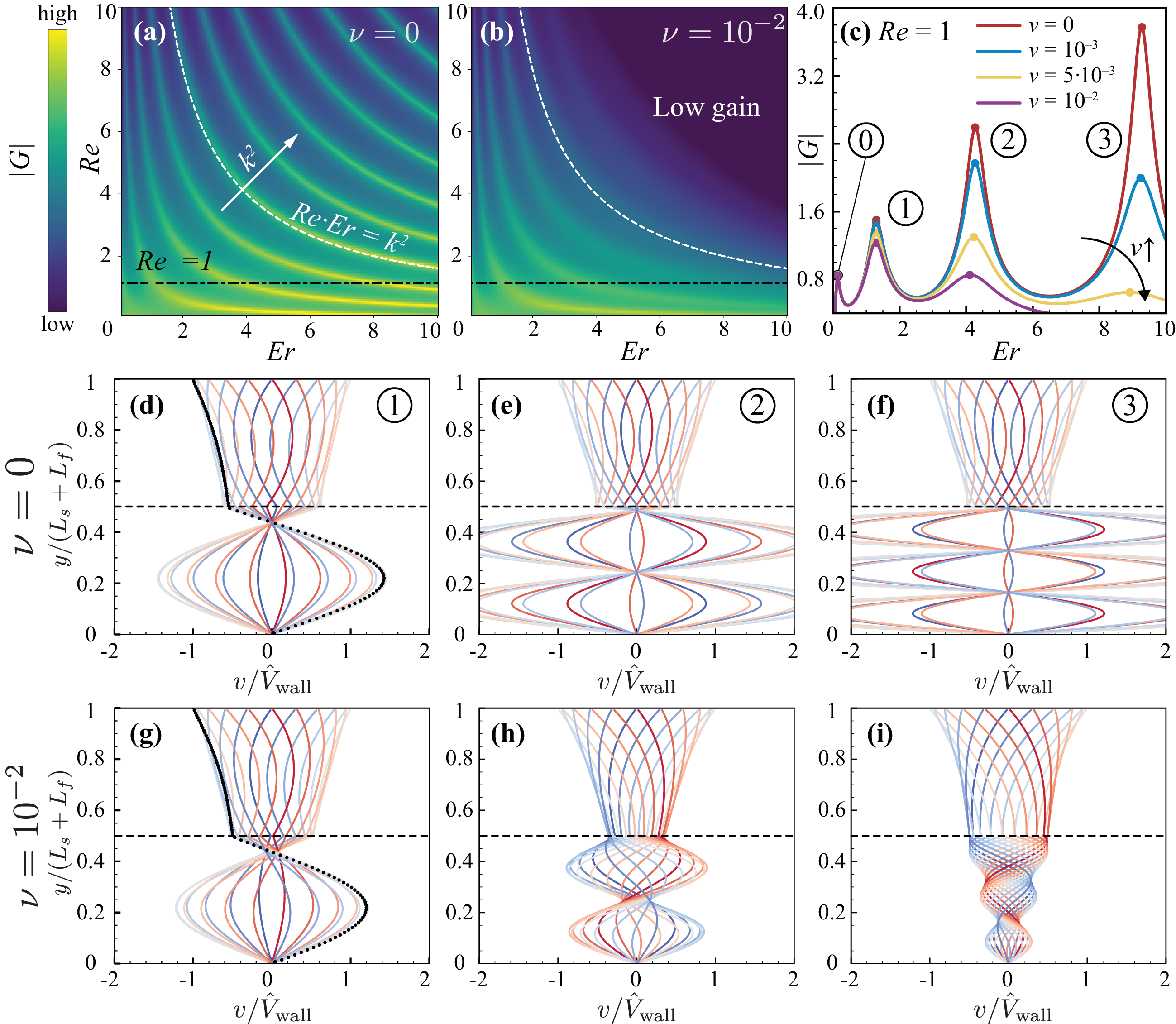}
\caption{\label{fig:gain_panel}High amplitude gains. Phase map of \( |G|\) as a function of \Rey,~\Er~shows distinct regions of high (bright yellow) and low (dark blue) gains in cases with \capsub{a} \( \nu = 0\), \capsub{b} \( \nu = 10^{-2} \), with maxima in both cases along \( \Rey \cdot \Er = k^2 \) (white dashed line), where \( k \) increases in the direction of the arrow in \capsub{a}. We fix \( \Rey = 1\) and plot \( |G|\) against \( \Er \) along the black, dashed line in \capsub{a} for different viscosity ratios \(\nu\) (colored) in \capsub{c}, which shows four distinct peaks \Circled{$0$} -- \Circled{$3$} at \( \Er = 0.14, 1.3, 4.27, 9.27\). Increasing viscosity decreases \( |G|\) peaks, especially for \Circled{$2$}, \Circled{$3$}. Plotting the velocity profiles corresponding to \Circled{$1$} -- \Circled{$3$} at \( \nu = 0\) in \capsub{d--f} and \( \nu = 10^{-2}\) in \capsub{g--i} reveals that they resemble harmonic standing waves within the solid.
% We then change \( \nu \) to \(10^{-2}\) keeping all other parameters fixed, and notice that |G| is reduced, as seen from the velocity profiles in \capsub{g--i}.
We confirm that these high gain results are indeed physical by plotting equivalent 2D DNS results as black scatter points in \capsub{d,g} where we notice agreement between the curves.}
\end{figure}
\par
In~\cref{fig:gain_panel}a, we plot \( |G| \) as a function of \( \Er, \Rey \) with \( \nu = 0\). As can be seen, characteristic gain peaks (\(|G| > 1\), bright yellow) take place and manifest as families of hyperbolae \( \Rey \cdot \Er = k^2 \). Here \( k = \frac{\pi}{\lambda}\) corresponds to discrete harmonic wavenumbers with wavelength \( \lambda = \left( \pi^2 \Rey \Er \right)^{-0.5} \), from~\cref{tab:params}. Hence, higher \( k \) corresponds to higher harmonics. To further gain intuition, we fix \( \Rey = 1\) (chosen because of its dynamic richness, see~\cref{fig:neo_hookean_sweep}), and plot \( |G| \) versus \( \Er \) to obtain the red curve of~\cref{fig:gain_panel}c. We see four distinct high-gain peaks \Circled{$0$} -- \Circled{$3$} of increasing amplitude, where the numbers represent the standing wave mode. In the cases \Circled{$1$} -- \Circled{$3$}, the solid displaces more than the driving wall~(\cref{fig:gain_panel}d--f), with velocity profiles corresponding to the first three harmonics.
Case \Circled{$0$} is characterized by the fact that maximal amplitudes occur at the interface, and not in the bulk, thus behaving as a standing wave with a free end at the interface.
% The symmetry between \( \Rey, \Er \) in accessing these harmonics
\par
Realistic materials include internal dissipation effects, which we enable here by adding viscosity to the solid. This drastically reduces \( |G| \) amplitudes, but preserves the hyperbolic structure of peaks, as seen in~\cref{fig:gain_panel}b.
\par
Lastly, 2D DNS simulations for \( \nu = 0, 10^{-2}\) (\cref{fig:gain_panel}d,f, black scatter points) validate our model predictions. Since it is numerically challenging to capture these high gain regimes, we propose \( \Rey = 1, \Er = 1.3, \nu = {0, 10^{-2}}\) for benchmarking numerical simulations, in addition to the parameter sets presented in~\cref{sec:neo_hookean_dynamics}.
\par
We have thus provided analytical solutions for the dynamics of a viscoelastic neo-Hookean
solid immersed in an oscillatory Couette flow system. We have derived a general solution to account
for arbitrary solid densities and viscosities in our setup, using two approaches---one in modal space generalizing the previous work of~\citet{sugiyama2011full}, and one in physical space. As a special limiting case, we recover the original solution of \citet{sugiyama2011full}
for a density-matched solid with zero viscosity.
Additionally, we recover analytical solutions of single~\cite{landau1987theoretical} and multi-fluid~\cite{sim2006stratified,leclaire2014unsteady}
Stokes--Couette flows in the limit of zero solid elasticity. Further, our solutions compare well against DNS results (\cref{fig:neo_hookean_simulations}). They are found to exhibit a range of behaviors (\cref{fig:neo_hookean_sweep}), including high gains~(\cref{fig:gain_panel}), with potential applications in biophysics and engineering. Next, we discuss the case of a generalized Mooney--Rivlin solid, which presents higher order non-linear effects within the solid.
\section{Generalization to Mooney--Rivlin solids}
\label{sec:mr_solution}
\subsection{Modal semi-analytical solutions}
In the case of a generalized Mooney--Rivlin solid, characterized by \(c_{3} \neq 0\), the hyperelastic stress
is proportional to the cubic power of strain~(\cref{eqn:stress}) which signifies a higher order
non-linear response to deformations. The resulting equations, whose non-linearity is overall captured via the parameter \(c = {c_{3}} / {c_{1}}\), resist closed-form analytical solutions.
% In these equations, we characterize the non-linearity via a parameter \(c = {c_{3}} / {c_{1}}\).
% To investigate the system response in this setting, we first conduct an asymptotic analysis for small non-linearity parameter \( c \) to uncover the solution structure, contrast it with the neo-Hookean solutions~(\cref{sec:phys_solution}), and gain intuition into parameter variations.
% We then translate our understanding
Then, to investigate the system response in this setting, we derive a semi-analytical solution using the Fourier series machinery of~\cref{sec:fourier_series}.

The solution strategy here is to employ a Fourier pseudospectral collocation scheme~\cite{sugiyama2011full}
for evaluating the nonlinear stress terms \(\sigma_{\mathrm{NL}, k}\) in
the governing~\cref{eqn:mom_solid_modal}, at a finite set of grid points
 \(x_j = (j + \frac{1}{2}) \Delta x\), with \(\Delta x= {L_s}/{K}\). All
other terms are treated as described in \cref{sec:fourier_series}.
% \ichang{Our choice of employing Fourier pseudo-spectral scheme here, as opposed to other discretization schemes, stems from the natural decomposition of the solution into sinusoidal modes (\cref{sec:fourier_series}) which enables faster convergence with the number of degrees of freedom \( K\)}.

Armed with this spatial discretization, we employ a numerical time
integration scheme to evolve the non-linear~\cref{eqn:mom_fluid_modal,eqn:mom_solid_modal}. Here, we use a second order constant timestepper (of timestep \( \Delta t\)) comprised of mixed Crank-Nicolson (implicit, for stability in the viscous updates)
and explicit Nyström (midpoint) rule for the second-order time derivatives~\cite{hairer1991solving}.
If we denote the \(n^{\textrm{th}}\) time level \(t = n \Delta t\) by a
superscript \(n\), then the prescribed wall velocity takes the analytical form
\begin{equation}
    V_{\textrm{wall}}^{(n+1)} := V_{\textrm{wall}}((n+1)(\Delta t)) =
    \imag{\hat{V}_{\textrm{wall}} \exp (i \omega \left( (n+1)\Delta t \right) )}.
\end{equation}
For the interface displacement \(U_{I}\) and fluid velocity update in~\cref{eqn:mom_fluid_modal}, we use the Crank-Nicolson scheme~\cite{hairer1991solving}
\begin{equation}
    \label{eq:displacement_update_maintext}
    U_{I}^{(n+1)} \approx U_{I}^{(n)}+\frac{\Delta t}{2}\left(V_{I}^{(n+1)}+V_{I}^{(n)}\right) + \order{\Delta t^2}
\end{equation}
and for updating the interface velocity \(V_{I}\) and solid displacements in~\cref{eqn:mom_solid_modal}, we utilize the explicit Nyström (midpoint) rule %discretizations at the \(n^{\textrm{th}}\) time step
% \begin{equation}
% \label{eqn:mom_solid_modal}
% -\frac{2(-1)^{k}}{\pi k} \frac{\mathrm{d} V_{I}}{\mathrm{d} t}+\frac{\mathrm{d}^{2} u_{s, k}}{\mathrm{d} t^{2}} + \nu_s\left(\frac{\pi k}{L_{s}}\right)^{2} \frac{\mathrm{d} u_{s, k}}{\mathrm{d} t} + \frac{2c_{1}}{\rho_s}\left(\frac{\pi k}{L_{s}}\right)^{2} u_{s, k}+\frac{\pi k}{\rho_s L_{s}} \sigma_{\mathrm{NL}, k}=0
% \end{equation}
%
\[ \left( \frac{\mathrm{d} V_{I}}{\mathrm{d} t} \right)^{(n)} \approx \frac{ V_{I}^{(n+1)} - V_{I}^{(n-1)} }{2 \Delta t} + \order{\Delta t^2}. \]
% \[ \left( \frac{\mathrm{d} u_{s, k}}{\mathrm{d} t} \right)^{(n)} \approx \frac{ u_{s, k}^{(n+1)} - u_{s, k}^{(n-1)} }{2 \Delta t} + \order{\Delta t^2} \]
% \[
% \left( \frac{\mathrm{d}^{2} u_{s, k}}{\mathrm{d} t^{2}}\right)^{(n)}  \approx \frac{u_{s, k}^{(n+1)} - 2 u_{s, k}^{(n)} + u_{s, k}^{(n-1)}}{(\Delta t)^2} + \order{\Delta t^2}.
% \]
%
Upon substituting these discretizations in the governing equations~\cref{eqn:mom_fluid_modal,eqn:mom_solid_modal},
and by invoking the modal stress balance of~\cref{eqn:stress_continuity_modal} at
every step, we obtain the solution, after standard (but tedious) algebraic manipulations. For brevity, we omit derivation details, which can be found in~\appref{app:mr_derivation}.
% This concludes the derivation of semi-analytical solutions for the case of a generalized Mooney--Rivlin solid material. We proceed to first validate our semi-analytical velocity profiles against those obtained from direct numerical simulations, \ichang{before relating it to the asymptotic results of~\cref{sec:asymptotics}.}
%
\subsection{Analysis of system behavior}\label{sec:mr_behavior}
\begin{figure}
\centering
\includegraphics[width=\textwidth]{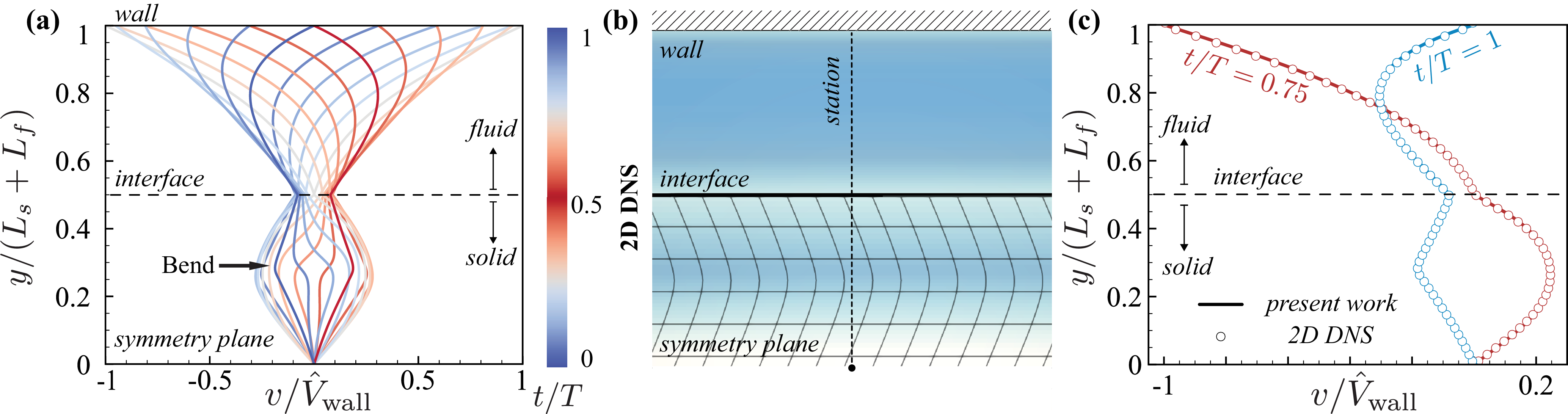}
\caption{
    \label{fig:mr_simulations}
    Comparison against simulations.
    \capsub{a} Non-dimensional semi-analytical velocity profiles in \(y\), for a visco-elastic solid with a generalized Mooney--Rivlin model. The system response is shown only in the upper half plane.  This system is characterized by \(Re = 2, \Er = 1, \nu = 0.1, c = c_{3} / c_{1} = 4, \delta_f = 0.4, \delta_s = 0.126, \lambda = 0.225 \). Additional details can be found in~\appref{app:fig_details}. Colors indicate \( t / T\). We also mark a high curvature bend in the profile with a black arrow. \capsub{b} We solve the above problem through 2D DNS~\cite{bhosale2021remeshed}
     %and run our simulations until the system reaches dynamical steady state ($t/T = 10$), and then sample quantities within the last cycle.
    with corresponding simulation parameters reported in~\appref{app:fig_details}. In this image, we mark the \(x\)-velocity field (orange/blue represent positive/negative velocity) and  deformation contours within the solid, with the interface marked (black, thick solid) for visual clarity. Upon plotting the velocity profiles at the highlighted station (black, dashed) in the center of the domain, we see good agreement with our analytical results across all times as shown in \capsub{c}. For the sake of clarity, we only show profiles at two different time instances. Here numerical results are plotted with scatter points whereas analytical results are plotted with a solid line.
}
\end{figure}

% \subsection{Generalized Mooney--Rivlin solid: Verification with direct numerical simulations}\label{sec:mooney_rivlin_verification}
% We start by highlighting a typical temporal non-dimensional velocity profile in the upper half plane obtained by solving the semi-analytical equations. This is shown in \cref{fig:mr_simulations}(a), for a system characterized by \(Re = 2, \Er = 1, \nu = 0.1, \delta_f = 0.4, \delta_s = 0.126, \lambda = 0.225\). We define an additional parameter \(c = {c_{3}} / {c_{1}}\) which captures the extent of higher-order non-linearity in the solid. Here we set \( c\) to an intermediate value of \(4\), chosen for demonstrating the non-linear behavior of the solid phase without making it too stiff.
We first validate our semi-analytical solutions against direct numerical simulations (\cref{fig:mr_simulations}) in the same setup of~\cref{fig:neo_hookean_simulations}, but with \( c = 4\) instead of zero. The choice of \( c \) is consistent with established biological tissue models~\cite{raghavan2000toward}. As can be seen in~\cref{fig:mr_simulations}a, the solid velocity profiles exhibit characteristics high-curvature bends (marked), differently from the neo-Hookean case (\cref{fig:neo_hookean_simulations}) on account of the additional material non-linearity.
% The resulting velocity profiles are shown in~\cref{fig:mr_simulations}a where we see that solid non-linearities manifest as high-curvature bends (marked).
% For velocity profiles with this value of \( c \), the non-linearity is clearly visible in the solid phase, where the solution now displays high curvature bends, marked in \cref{fig:mr_simulations}(a). Formally, these bends are waves arising from the contribution of higher order non-linearities to high frequency Fourier modes in our semi-analytical solution,~\ichang{similar to our asymptotic solution}. We confirm this by noting that these bends are absent in the case of an equivalent neo-Hookean solid without non-linearities~(\cref{fig:neo_hookean_simulations}(a)), for the same \(Re, \Er, \nu \) parameters.
% Comparison of these profiles with results from simulations (\cref{fig:mr_simulations}b) indicate a good match (\cref{fig:mr_simulations}c).
% Next, we compare these results against simulations, whose setup and solid deformation are shown in~\cref{fig:mr_simulations}(b). Here, consistently, the same bends are reflected in the solid deformation contours, which quantitatively agree with our semi-analytic solution, at multiple temporal instants~(\cref{fig:mr_simulations}(c)).
Further, as illustrated in~\cref{fig:mr_simulations}c, our semi-analytical solutions are found to agree well with direct numerical simulations.
\begin{figure}
\centering
\includegraphics[width=1.0\textwidth]{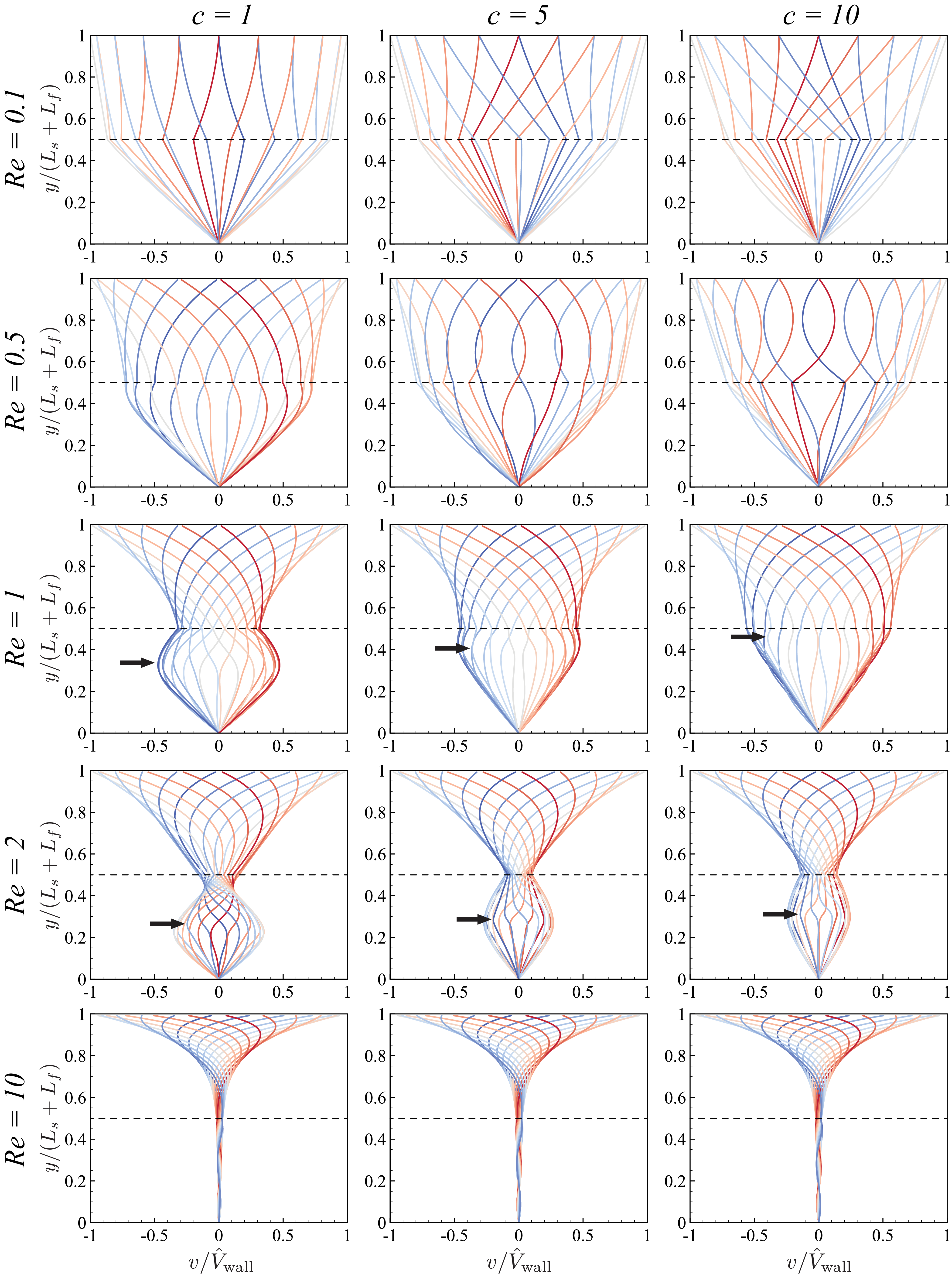}
\caption{\label{fig:mr_sweep}Dynamics with parametric variation. Non-dimensional velocity profile in \(y\) for parametric changes in \( \Rey = [0.1, 0.5, 1, 2, 10] \; \times \; c = [1, 5, 10]\) with \( \Er = 1, \nu = 0.1, \rho = 1\). The system response is shown only in the upper half plane. The black, dashed lines indicate the solid--fluid interface. Colors represent \(t / T\), with the same colorbar as \cref{fig:linear_velocity}. We also mark the location of prominent high-curvature bends in the profiles with a black arrow.}
\end{figure}
\par
Next, in the spirit of~\cref{fig:neo_hookean_sweep}, we highlight system responses upon varying both degree of solid non-linearity (\( c\)) and viscosity (\(\Rey\)). Throughout this exploration, we fix the elastic to viscous contributions by setting \( Er = 1\) and \( \nu = 0.1\). These values are informed by the rich dynamics of the central column of~\cref{fig:neo_hookean_sweep}. We then span \( c = [1, 5, 10] \; \times \; \Rey = [0.1, 0.5, 1, 2, 10]\) and report responses of the system in~\cref{fig:mr_sweep}.
% . The range for the non-linearity parameter \( c \) reflects biologically relevant values found in aortic tissues~\cite{raghavan2000toward}.
%in patients with aneurysms, and accounts for patient-specific measurements and experimental errors.
% Additionally, we span an order of magnitude in \( c \) to showcase material responses, from limited to strongly non-linear characteristics. We depict these responses in~\cref{fig:mr_sweep}, with parametric details found in the image caption. % We also mark the locations of bends in these profiles, along with the length scales \( \delta_f, \delta_s, \lambda\).

% We first discuss the effects of non-linear elasticity \(c\) variation.
For solids with small \(c\), we expect dynamics similar to the neo-Hookean counterpart, as the visco-elastic response is linear to a first order of approximation. This is confirmed from the solid zone profiles in the left column.
% where the bends from non-linearity are less prominent.
Increasing the non-linearity coefficient stiffens the solid, constraining deformation velocities (narrower envelopes) as well as producing sharper bends (marked), as we move from left to right in~\cref{fig:mr_sweep}.
%reflects
% expect an effective stiffening of the solid due to increasing elastic stress contributions from the non-linear term. This non-linear contribution is clearly visible at high \( \Rey > 1\) and

Changing viscosity (\Rey)~affects the response in a fashion similar to the neo-Hookean case (\cref{fig:neo_hookean_sweep}), where profile curvatures in both fluid and solid phases progressively get concentrated within sharper boundary layers, as we move from top to bottom in~\cref{fig:mr_sweep}.
% We note that the bends in the Mooney--Rivlin solid contribute to a local increase in curvature. This is why we see the effect of non-linearity more prominently at higher \Rey~where the boundary layers are typically thinner.
%
\par
\begin{figure}
\centering
\includegraphics[width=1.0\textwidth]{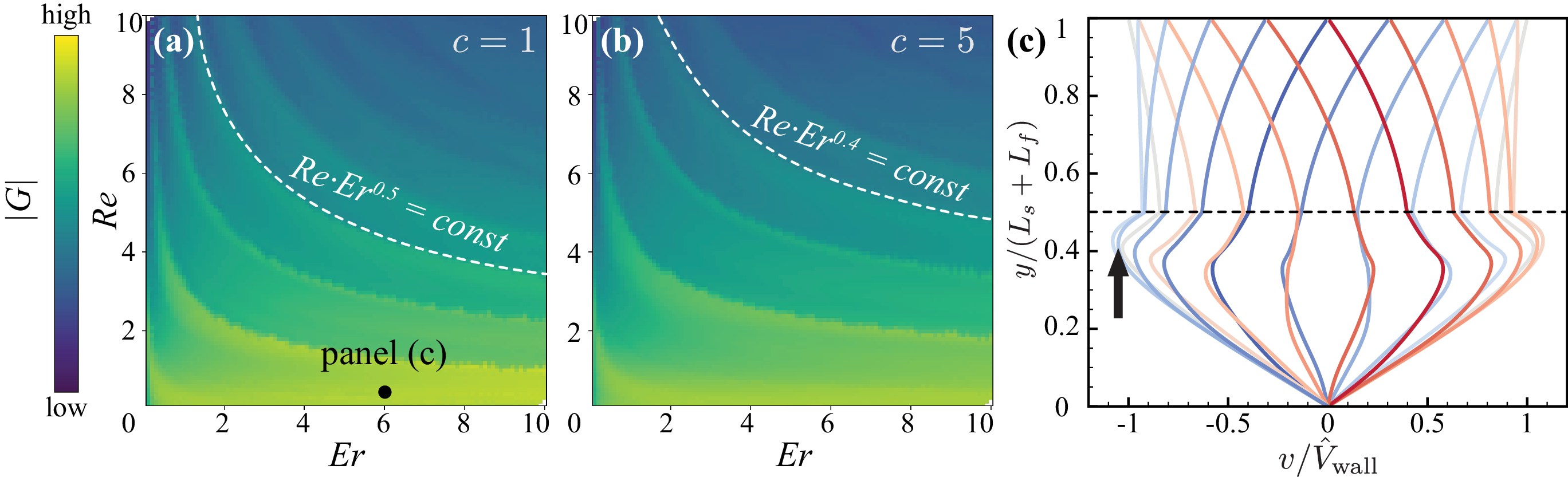}
\caption{\label{fig:mr_gains}High amplitude gains for a Mooney--Rivlin solid. Phase map of numerically measured \( |G|\) as a function of \Rey,~\Er~for \( \nu = 10^{-3}\) and
\capsub{a} \( c = 1\), \capsub{b} \( c = 5 \) shows regions of high (bright yellow) and low (dark blue) gains with maxima along \( \Rey \cdot \Er^{0.5} = \textrm{const}\) and \( \Rey \cdot \Er^{0.4} = \textrm{const}\) respectively, deviating from the hyperbolae seen in the neo-Hookean case. Velocity profiles of a representative case with \( \Rey = 0.2, \Er = 6, c = 1 \) marked in \capsub{a} and plotted in \capsub{c} showcases high-gain bends (marked) characteristic of strong nonlinear effects in the solid.}
\end{figure}

Finally, we investigate whether the high gain regimes seen in~\cref{sec:gains} exist for Mooney--Rivlin  solids, and if so, under what conditions. Here, unlike~\cref{sec:gains}, there is no mathematical guidance to identify high-gain parameters, thus we explore the phase space \( \Er-\Rey \) numerically for the representative cases \( c = 1 \) (\cref{fig:mr_gains}a) and \( c = 5\) (\cref{fig:mr_gains}b). For \( c =1\), we see in \cref{fig:mr_gains}a that high-gain peaks (bright yellow), although less pronounced, still occur in a regular structure that departs from the hyperbolae seen in the neo-Hookean case, and instead lie on the  curve \( \Rey \cdot \Er^{0.5} = \textrm{constant} \). In~\cref{fig:mr_gains}c, we report the velocity profiles of a representative high gain case, and note that \( |v / \hat{V}_{\textrm{wall}}| \) hardly exceeds 1, as opposed to the neo-Hookean cases of~\cref{fig:gain_panel}. As we increase \( c \) from \( 1\) to \(5\), we observe that the peaks spread apart and gains further diminish (\cref{fig:mr_gains}b).
We conclude that the cubic non-linear term that characterize Mooney--Rivlin solids locally stiffens the material, reducing its propensity to deform and shear.
% We then depict the velocity profile of one such high-gain case in ~\cref{fig:mr_gains}c, and see that high gains can be attributed to bends in the velocity profiles, in addition to the harmonic structure seen in~\cref{sec:gains}. Lastly, an increase in \(c\) from 1 (in ~\cref{fig:mr_gains}a) to 5 (in ~\cref{fig:mr_gains}b) alters the number of high-gain peaks and presents diminished overall gains.
%
\section{Conclusion}\label{sec:conc}
We have presented solutions for an oscillatory Coutte setup involving parallel
viscoelastic solid--fluid layers sandwiched between two oscillating walls. We are motivated
by the paucity of minimal yet representative (hyper-)elastohydrodynamic systems
that can be analytically and rigorously analysed, given their relevance and
ubiquity in biophysical and engineering settings.
Here we consider visco-hyperelastic solids with arbitrary density and viscosity immersed in a Newtonian fluid.
% made of neo-Hookean and generalized Mooney--Rivlin materials, which we sequentially analyzed.
% and analyzed them case by case.
% Here, we generalize the analysis of~\citet{sugiyama2011full} to account for solid hyperelasticity, viscosity and density mismatch.
First, for a sandwiched viscous neo-Hookean solid, the governing equations
simplify to an analytically
tractable problem for which we derive two equivalent solutions. One is obtained by analytically solving the homogeneous Helmholtz equations derived from the governing PDEs. The other
is based on partitioned Fourier-series
expansions, providing a general machinery that can be applied to different material models. Obtained solutions, in the limit of zero elasticity, recover the classical Stokes--Couette solutions for single and
multiple, immiscible fluids.
% We then consider a generalized Mooney--Rivlin material.
% with internal viscosity, the simplified equations retain characteristic non-linearity, and their solution resists a closed-form description. %Here we turn to (semi-)numerical
% \ichang{For small non-linearities, we first derive an asymptotic solution based on the viscous neo-Hookean Helmholtz solutions above. For large non-linearities, the asymptotic solution breaks down and hence we
Semi-analytic solutions for generalized Mooney--Rivlin materials are also derived, and numerically solved using a pseudospectral scheme, based on the partitioned Fourier-series expansions introduced in the neo-Hookean case. For both neo-Hookean and Mooney--Rivlin materials, we report quantitative agreement with direct numerical flow--structure interaction simulations, and further explore system behavior upon parametric changes in solid elasticity, fluid
viscosity and higher order elastic nonlinearities. %in the case of a generalized Mooney--Rivlin solid.
This analysis allows us to identify the spatio-temporal scales at play, assess the degree of flow--structure coupling, and highlight differences between neo-Hookean and Mooney--Rivlin models. We also
highlight regimes of high solid phase displacements, which result from standing wave harmonics. The proposed hyperelastic oscillatory Couette system, and its analysis, can find application in a range of biophysical settings, from in-situ bio-film formation and
synovial joint mechanics to solid-rheological characterization. Furthermore, the proposed setup may well serve as a
benchmark to rigorously test numerical solvers for coupled fluid--elastic solid
interactions.  Finally, for building fundamental fluid mechanics intuition, we provide, as free software, an interactive, online sandbox demonstrating our results~(\appref{app:sandbox}),
together with our computational code.

\section{Declaration of Interests}
The authors report no conflict of interest.

\bibliographystyle{jfm}
\bibliography{parallel_slab_lit.bib}

\begin{thebibliography}{43}
\expandafter\ifx\csname natexlab\endcsname\relax\def\natexlab#1{#1}\fi
\def\au#1{#1} \def\ed#1{#1} \def\yr#1{#1}\def\at#1{#1}\def\jt#1{\textit{#1}}
  \def\bt#1{#1}\def\bvol#1{\textbf{#1}} \def\vol#1{#1} \def\pg#1{#1}
  \def\publ#1{#1}\def\arxiv#1{#1}\def\org#1{#1}\def\st#1{\textit{#1}}

\bibitem[Alben {\em et~al.\/}(2002)Alben, Shelley \& Zhang]{alben2002drag}
{\sc \au{Alben, Silas}, \au{Shelley, Michael} \& \au{Zhang, Jun}} \yr{2002}
  \at{Drag reduction through self-similar bending of a flexible body}.
  \jt{Nature}  \bvol{420}~(6915),  \pg{479--481}.

\bibitem[Alben {\em et~al.\/}(2004)Alben, Shelley \&
  Zhang]{alben2004flexibility}
{\sc \au{Alben, Silas}, \au{Shelley, Michael} \& \au{Zhang, Jun}} \yr{2004}
  \at{How flexibility induces streamlining in a two-dimensional flow}.
  \jt{Physics of Fluids}  \bvol{16}~(5),  \pg{1694--1713}.

\bibitem[Argentina \& Mahadevan(2005)]{argentina2005fluid}
{\sc \au{Argentina, M{\'e}d{\'e}ric} \& \au{Mahadevan, L}} \yr{2005}
  \at{Fluid-flow-induced flutter of a flag}.  \jt{Proceedings of the National
  Academy of Sciences}  \bvol{102}~(6),  \pg{1829--1834}.

\bibitem[Argentina {\em et~al.\/}(2007)Argentina, Skotheim \&
  Mahadevan]{argentina2007settling}
{\sc \au{Argentina, Mederic}, \au{Skotheim, J} \& \au{Mahadevan, L}} \yr{2007}
  \at{Settling and swimming of flexible fluid-lubricated foils}.  \jt{Physical
  review letters}  \bvol{99}~(22),  \pg{224503}.

\bibitem[Bakker {\em et~al.\/}(2003)Bakker, Huijs, de~Vries, Klijnstra,
  Busscher \& van~der Mei]{bakker2003bacterial}
{\sc \au{Bakker, Dewi~P}, \au{Huijs, Frank~M}, \au{de~Vries, Joop},
  \au{Klijnstra, Job~W}, \au{Busscher, Henk~J} \& \au{van~der Mei, Henny~C}}
  \yr{2003}  \at{Bacterial deposition to fluoridated and non-fluoridated
  polyurethane coatings with different elastic modulus and surface tension in a
  parallel plate and a stagnation point flow chamber}.  \jt{Colloids and
  Surfaces B: Biointerfaces}  \bvol{32}~(3),  \pg{179--190}.

\bibitem[Barthes-Biesel(2016)]{barthes2016motion}
{\sc \au{Barthes-Biesel, Dominique}} \yr{2016}  \at{Motion and deformation of
  elastic capsules and vesicles in flow}.  \jt{Annual Review of fluid
  mechanics}  \bvol{48},  \pg{25--52}.

\bibitem[Bhosale {\em et~al.\/}(2020)Bhosale, Esmaili, Bhar \&
  Jung]{bhosale2020bending}
{\sc \au{Bhosale, Yashraj}, \au{Esmaili, Ehsan}, \au{Bhar, Kinjal} \& \au{Jung,
  Sunghwan}} \yr{2020}  \at{Bending, twisting and flapping leaf upon raindrop
  impact}.  \jt{Bioinspiration \& Biomimetics}  \bvol{15}~(3),  \pg{036007}.

\bibitem[Bhosale {\em et~al.\/}(2021)Bhosale, Parthasarathy \&
  Gazzola]{bhosale2021remeshed}
{\sc \au{Bhosale, Yashraj}, \au{Parthasarathy, Tejaswin} \& \au{Gazzola,
  Mattia}} \yr{2021}  \at{A remeshed vortex method for mixed rigid/soft body
  fluid--structure interaction}.  \jt{Journal of Computational Physics}  \pg{p.
  110577}.

\bibitem[Bodn{\'a}r {\em et~al.\/}(2014)Bodn{\'a}r, Galdi \&
  Ne{\v{c}}asov{\'a}]{bodnar2014fluid}
{\sc \au{Bodn{\'a}r, Tom{\'a}{\v{s}}}, \au{Galdi, Giovanni~P} \&
  \au{Ne{\v{c}}asov{\'a}, {\v{S}}{\'a}rka}} \yr{2014} {\em Fluid-structure
  interaction and biomedical applications\/}.  \publ{Springer}.

\bibitem[Bower(2009)]{bower2009applied}
{\sc \au{Bower, Allan~F}} \yr{2009} {\em Applied mechanics of solids\/}.
  \publ{CRC press}.

\bibitem[Carr {\em et~al.\/}(1976)Carr, Shen \& Hermans]{carr1976physical}
{\sc \au{Carr, Marcus~E}, \au{Shen, Linus~L} \& \au{Hermans, Jan}} \yr{1976}
  \at{A physical standard of fibrinogen: measurement of the elastic modulus of
  dilute fibrin gels with a new elastometer}.  \jt{Analytical biochemistry}
  \bvol{72}~(1-2),  \pg{202--211}.

\bibitem[Christov(2021)]{christov2021soft}
{\sc \au{Christov, Ivan~C}} \yr{2021}  \at{Soft hydraulics: from newtonian to
  complex fluid flows through compliant conduits}.  \jt{arXiv preprint
  arXiv:2106.07164} .

\bibitem[Di~Carlo(2009)]{di2009inertial}
{\sc \au{Di~Carlo, Dino}} \yr{2009}  \at{Inertial microfluidics}.  \jt{Lab on a
  Chip}  \bvol{9}~(21),  \pg{3038--3046}.

\bibitem[Dowell \& Hall(2001)]{dowell2001modeling}
{\sc \au{Dowell, Earl~H} \& \au{Hall, Kenneth~C}} \yr{2001}  \at{Modeling of
  fluid-structure interaction}.  \jt{Annual review of fluid mechanics}
  \bvol{33}~(1),  \pg{445--490}.

\bibitem[Dowson \& Jin(1986)]{dowson1986micro}
{\sc \au{Dowson, D} \& \au{Jin, Zhong-Min}} \yr{1986}
  \at{Micro-elastohydrodynamic lubrication of synovial joints}.
  \jt{Engineering in medicine}  \bvol{15}~(2),  \pg{63--65}.

\bibitem[Duncombe {\em et~al.\/}(2015)Duncombe, Tentori \&
  Herr]{duncombe2015microfluidics}
{\sc \au{Duncombe, Todd~A}, \au{Tentori, Augusto~M} \& \au{Herr, Amy~E}}
  \yr{2015}  \at{Microfluidics: reframing biological enquiry}.  \jt{Nature
  Reviews Molecular Cell Biology}  \bvol{16}~(9),  \pg{554--567}.

\bibitem[Gazzola {\em et~al.\/}(2015)Gazzola, Argentina \&
  Mahadevan]{gazzola2015gait}
{\sc \au{Gazzola, Mattia}, \au{Argentina, M{\'e}d{\'e}ric} \& \au{Mahadevan,
  Lakshminarayanan}} \yr{2015}  \at{Gait and speed selection in slender
  inertial swimmers}.  \jt{Proceedings of the National Academy of Sciences}
  \bvol{112}~(13),  \pg{3874--3879}.

\bibitem[Grotberg \& Jensen(2004)]{grotberg2004biofluid}
{\sc \au{Grotberg, James~B} \& \au{Jensen, Oliver~E}} \yr{2004}  \at{Biofluid
  mechanics in flexible tubes}.  \jt{Annu. Rev. Fluid Mech.}  \bvol{36},
  \pg{121--147}.

\bibitem[Guimar{\~a}es {\em et~al.\/}(2020)Guimar{\~a}es, Gasperini, Marques \&
  Reis]{guimaraes2020stiffness}
{\sc \au{Guimar{\~a}es, Carlos~F}, \au{Gasperini, Luca}, \au{Marques,
  Alexandra~P} \& \au{Reis, Rui~L}} \yr{2020}  \at{The stiffness of living
  tissues and its implications for tissue engineering}.  \jt{Nature Reviews
  Materials}  \bvol{5}~(5),  \pg{351--370}.

\bibitem[Hairer {\em et~al.\/}(1991)Hairer, N{\o}rsett \&
  Wanner]{hairer1991solving}
{\sc \au{Hairer, Ernst}, \au{N{\o}rsett, Syvert~P} \& \au{Wanner, Gerhard}}
  \yr{1991} {\em Solving ordinary differential equations I, Nonstiff
  problems\/}.  \publ{Springer-Vlg}.

\bibitem[Gad-el Hak(2002)]{gad2002compliant}
{\sc \au{Gad-el Hak, Mohamed}} \yr{2002}  \at{Compliant coatings for drag
  reduction}.  \jt{Progress in Aerospace Sciences}  \bvol{38}~(1),
  \pg{77--99}.

\bibitem[Heil \& Hazel(2011)]{heil2011fluid}
{\sc \au{Heil, Matthias} \& \au{Hazel, Andrew~L}} \yr{2011}
  \at{Fluid-structure interaction in internal physiological flows}.  \jt{Annual
  review of fluid mechanics}  \bvol{43},  \pg{141--162}.

\bibitem[Heil {\em et~al.\/}(2008)Heil, Hazel \& Smith]{heil2008mechanics}
{\sc \au{Heil, Matthias}, \au{Hazel, Andrew~L} \& \au{Smith, Jaclyn~A}}
  \yr{2008}  \at{The mechanics of airway closure}.  \jt{Respiratory physiology
  \& neurobiology}  \bvol{163}~(1-3),  \pg{214--221}.

\bibitem[Kou {\em et~al.\/}(2017)Kou, Pandolfino, Kahrilas \&
  Patankar]{kou2017simulation}
{\sc \au{Kou, Wenjun}, \au{Pandolfino, John~E}, \au{Kahrilas, Peter~J} \&
  \au{Patankar, Neelesh~A}} \yr{2017}  \at{Simulation studies of the role of
  esophageal mucosa in bolus transport}.  \jt{Biomechanics and modeling in
  mechanobiology}  \bvol{16}~(3),  \pg{1001--1009}.

\bibitem[Landau \& Lifshitz(1987)]{landau1987theoretical}
{\sc \au{Landau, LD} \& \au{Lifshitz, EM}} \yr{1987} Theoretical physics, vol.
  6, fluid mechanics.

\bibitem[Leclaire {\em et~al.\/}(2014)Leclaire, Pellerin, Reggio \&
  Tr{\'e}panier]{leclaire2014unsteady}
{\sc \au{Leclaire, S}, \au{Pellerin, N}, \au{Reggio, M} \& \au{Tr{\'e}panier,
  JY}} \yr{2014}  \at{Unsteady immiscible multiphase flow validation of a
  multiple-relaxation-time lattice boltzmann method}.  \jt{Journal of Physics
  A: Mathematical and Theoretical}  \bvol{47}~(10),  \pg{105501}.

\bibitem[Li {\em et~al.\/}(2013)Li, Vlahovska \& Karniadakis]{li2013continuum}
{\sc \au{Li, Xuejin}, \au{Vlahovska, Petia~M} \& \au{Karniadakis, George~Em}}
  \yr{2013}  \at{Continuum-and particle-based modeling of shapes and dynamics
  of red blood cells in health and disease}.  \jt{Soft matter}  \bvol{9}~(1),
  \pg{28--37}.

\bibitem[Mihai \& Goriely(2017)]{mihai2017characterize}
{\sc \au{Mihai, L~Angela} \& \au{Goriely, Alain}} \yr{2017}  \at{How to
  characterize a nonlinear elastic material? a review on nonlinear constitutive
  parameters in isotropic finite elasticity}.  \jt{Proceedings of the Royal
  Society A: Mathematical, Physical and Engineering Sciences}
  \bvol{473}~(2207),  \pg{20170607}.

\bibitem[Nalim {\em et~al.\/}(2004)Nalim, Pekkan, Sun \&
  Yokota]{nalim2004oscillating}
{\sc \au{Nalim, Razi}, \au{Pekkan, Kerem}, \au{Sun, Hui~Bin} \& \au{Yokota,
  Hiroki}} \yr{2004}  \at{Oscillating couette flow for in vitro cell loading}.
  \jt{Journal of biomechanics}  \bvol{37}~(6),  \pg{939--942}.

\bibitem[Pozrikidis(2003)]{pozrikidis2003modeling}
{\sc \au{Pozrikidis, Constantine}} \yr{2003} {\em Modeling and simulation of
  capsules and biological cells\/}.  \publ{CRC Press}.

\bibitem[Raghavan \& Vorp(2000)]{raghavan2000toward}
{\sc \au{Raghavan, ML} \& \au{Vorp, David~A}} \yr{2000}  \at{Toward a
  biomechanical tool to evaluate rupture potential of abdominal aortic
  aneurysm: identification of a finite strain constitutive model and evaluation
  of its applicability}.  \jt{Journal of biomechanics}  \bvol{33}~(4),
  \pg{475--482}.

\bibitem[Sim(2006)]{sim2006stratified}
{\sc \au{Sim, Woo-Gun}} \yr{2006}  \at{Stratified steady and unsteady two-phase
  flows between two parallel plates}.  \jt{Journal of mechanical science and
  technology}  \bvol{20}~(1),  \pg{125}.

\bibitem[Song {\em et~al.\/}(2015)Song, Koo \& Ren]{song2015effects}
{\sc \au{Song, Fangchao}, \au{Koo, Hyun} \& \au{Ren, Dacheng}} \yr{2015}
  \at{Effects of material properties on bacterial adhesion and biofilm
  formation}.  \jt{Journal of dental research}  \bvol{94}~(8),
  \pg{1027--1034}.

\bibitem[Sugiyama {\em et~al.\/}(2010)Sugiyama, Ii, Takeuchi, Takagi \&
  Matsumoto]{sugiyama2010full}
{\sc \au{Sugiyama, Kazuyasu}, \au{Ii, Satoshi}, \au{Takeuchi, Shintaro},
  \au{Takagi, Shu} \& \au{Matsumoto, Yoichiro}} \yr{2010}  \at{Full eulerian
  simulations of biconcave neo-hookean particles in a poiseuille flow}.
  \jt{Computational Mechanics}  \bvol{46}~(1),  \pg{147--157}.

\bibitem[Sugiyama {\em et~al.\/}(2011)Sugiyama, Ii, Takeuchi, Takagi \&
  Matsumoto]{sugiyama2011full}
{\sc \au{Sugiyama, Kazuyasu}, \au{Ii, Satoshi}, \au{Takeuchi, Shintaro},
  \au{Takagi, Shu} \& \au{Matsumoto, Yoichiro}} \yr{2011}  \at{A full eulerian
  finite difference approach for solving fluid--structure coupling problems}.
  \jt{Journal of Computational Physics}  \bvol{230}~(3),  \pg{596--627}.

\bibitem[Sun(2010)]{sun2010mechanical}
{\sc \au{Sun, Hui~B}} \yr{2010}  \at{Mechanical loading, cartilage degradation,
  and arthritis}.  \jt{Annals of the New York Academy of Sciences}
  \bvol{1211}~(1),  \pg{37--50}.

\bibitem[Sun {\em et~al.\/}(2003)Sun, Nalim \& Yokota]{sun2003expression}
{\sc \au{Sun, Hui~Bin}, \au{Nalim, Razi} \& \au{Yokota, Hiroki}} \yr{2003}
  \at{Expression and activities of matrix metalloproteinases under oscillatory
  shear in il-1-stimulated synovial cells}.  \jt{Connective tissue research}
  \bvol{44}~(1),  \pg{42--49}.

\bibitem[Tytell {\em et~al.\/}(2016)Tytell, Leftwich, Hsu, Griffith, Cohen,
  Smits, Hamlet \& Fauci]{tytell2016role}
{\sc \au{Tytell, Eric~D}, \au{Leftwich, Megan~C}, \au{Hsu, Chia-Yu},
  \au{Griffith, Boyce~E}, \au{Cohen, Avis~H}, \au{Smits, Alexander~J},
  \au{Hamlet, Christina} \& \au{Fauci, Lisa~J}} \yr{2016}  \at{Role of body
  stiffness in undulatory swimming: insights from robotic and computational
  models}.  \jt{Physical Review Fluids}  \bvol{1}~(7),  \pg{073202}.

\bibitem[Velve-Casquillas {\em et~al.\/}(2010)Velve-Casquillas, Le~Berre, Piel
  \& Tran]{velve2010microfluidic}
{\sc \au{Velve-Casquillas, Guilhem}, \au{Le~Berre, Ma{\"e}l}, \au{Piel,
  Matthieu} \& \au{Tran, Phong~T}} \yr{2010}  \at{Microfluidic tools for cell
  biological research}.  \jt{Nano today}  \bvol{5}~(1),  \pg{28--47}.

\bibitem[Vlahovska \& Gracia(2007)]{vlahovska2007dynamics}
{\sc \au{Vlahovska, Petia~M} \& \au{Gracia, Ruben~Serral}} \yr{2007}
  \at{Dynamics of a viscous vesicle in linear flows}.  \jt{Physical Review E}
  \bvol{75}~(1),  \pg{016313}.

\bibitem[Wang \& Christov(2019)]{wang2019theory}
{\sc \au{Wang, Xiaojia} \& \au{Christov, Ivan~C}} \yr{2019}  \at{Theory of the
  flow-induced deformation of shallow compliant microchannels with thick
  walls}.  \jt{Proceedings of the Royal Society A}  \bvol{475}~(2231),
  \pg{20190513}.

\bibitem[Wu {\em et~al.\/}(2018)Wu, Aroush, Asnacios, Chen, Dokukin, Doss,
  Durand-Smet, Ekpenyong, Guck, Guz {\em et~al.\/}]{wu2018comparison}
{\sc \au{Wu, Pei-Hsun}, \au{Aroush, Dikla Raz-Ben}, \au{Asnacios, Atef},
  \au{Chen, Wei-Chiang}, \au{Dokukin, Maxim~E}, \au{Doss, Bryant~L},
  \au{Durand-Smet, Pauline}, \au{Ekpenyong, Andrew}, \au{Guck, Jochen},
  \au{Guz, Nataliia~V} \& \au{others}} \yr{2018}  \at{A comparison of methods
  to assess cell mechanical properties}.  \jt{Nature methods}  \bvol{15},
  \pg{491--498}.

\bibitem[Zhu \& Jane~Wang(2011)]{dong2011elastohydrodynamic}
{\sc \au{Zhu, Dong} \& \au{Jane~Wang, Q.}} \yr{2011}  \at{Elastohydrodynamic
  lubrication: A gateway to interfacial mechanics—review and prospect}.
  \jt{Journal of Tribology}  \bvol{133}~(4).

\end{thebibliography}
\end{document}

\maketitle
\section{Details of neo-Hookean analytical solution} \label{app:neo_hookean_details}
Here we detail the analytical solution in case of a neo-Hookean solid. The solution is
\begin{equation*}
\begin{aligned}
\hat{v}_{f}(\tilde{y}) &= A \exp{\left( k_{f}\frac{\tilde{y}}{L_f} \right)} + B \exp{\left( -k_{f}\frac{\tilde{y}}{{L_f}} \right)} \quad &&\tilde{y} \in [0, L_f) \\
\hat{u}_{s}(y) &= C \exp{\left( k_{s}\frac{y}{L_s} \right)} + D \exp{\left( -k_{s}\frac{y}{L_s} \right)} \quad && y \in [0, L_s),
\end{aligned}
\end{equation*}
where we determine the coefficients \(A, B, C, D \) using the boundary~(\mainref{eqn:simplifed_bc}) and
interface conditions~(\mainref{eqn:stress_continuity}), which lead to the following system of equations
\begin{equation}
\begin{aligned}
\left(\begin{array}{cccc}
\mathrm{e}^{k_{f}} & \mathrm{e}^{-k_{f}} & 0 & 0 \\
1 & 1 & -i\omega\mathrm{e}^{k_{s}} & -i\omega\mathrm{e}^{-k_{s}} \\
1 & -1 & M_{33} & M_{34} \\
0 & 0 & 1 & 1
\end{array}\right) \cdot\left(\begin{array}{c}
A \\
B \\
C \\
D
\end{array}\right)=\left(\begin{array}{c}
\hat{V}_{\textrm{wall}} \\
0 \\
0 \\
0
\end{array}\right),
\end{aligned}
\end{equation}
with
\begin{equation}
\begin{aligned}
M_{33} &= -i\omega \frac{k_s}{k_f} \frac{L_f}{L_s} \left[\left(\frac{\nu_s}{\nu_f}\right) \left(\frac{\rho_s}{\rho_f}\right) - i \left(\frac{2c_1}{\omega \mu_f}\right)  \right] \exp\left( k_s\right)  \\
M_{34} &= +i\omega \frac{k_s}{k_f} \frac{L_f}{L_s} \left[\left(\frac{\nu_s}{\nu_f}\right) \left(\frac{\rho_s}{\rho_f}\right) - i \left(\frac{2c_1}{\omega \mu_f}\right)  \right] \exp \left( -k_s\right)
% \lambda_2 \exp \left( -\lambda_2 L_s \right) \left[ 2c_1 + i \omega \mu_s \right]
\end{aligned}
\end{equation}
which we can solve for to yield the final solutions. The resulting coefficients are
\begin{equation}
    \begin{aligned}
    k_{f} &= \frac{(1 + i)}{\sqrt{2}} \left( L_f^{-1} \left({\nu_f}/\omega\right)^{0.5} \right)^{-1} = \exp{\left(i\pi / 4\right)} \delta^{-1}_f \\ % = \exp{\left(i\pi / 4\right)} \left(\frac{\Rey}{\dot{\gamma}}\right)^{0.5} \\
    k_{s} &= \frac{i}{\sqrt{\left(\left(\omega L_s\right)^{-1}\left({2c_1}/{\rho_s}\right)^{0.5}\right)^2 + i \left(L^{-1}_s \left(\nu_s/\omega\right)^{0.5} \right)^2}} = {i}\left(\lambda^2 + i\delta^2_s\right)^{-0.5}\\
    \alpha &= \frac{L_f}{L_s} \frac{k_s}{k_f} \left( \rho \nu - i \frac{\dot{\gamma}}{\Er} \right) \\
    A &= \hat{V}_{\textrm{wall}} \frac{\left[ \left(e^{-k_s}\right) \left( 1 - \alpha \right) - \left(e^{+k_s}\right) \left( 1 + \alpha \right)\right]}{\left[ \left(e^{k_f-k_s} - e^{-(k_f-k_s)}\right) \left( 1 - \alpha \right) - \left(e^{k_f + k_s} - e^{-(k_f + k_s)}\right) \left( 1 + \alpha \right)\right]} \\
    B &= \hat{V}_{\textrm{wall}} \frac{\left[ \left(-e^{k_s}\right) \left( 1 - \alpha \right) + \left(e^{-k_s}\right) \left( 1 + \alpha \right)\right]}{\left[ \left(e^{k_f-k_s} - e^{-(k_f-k_s)}\right) \left( 1 - \alpha \right) - \left(e^{k_f + k_s} - e^{-(k_f + k_s)}\right) \left( 1 + \alpha \right)\right]} \\
    C &= \frac{-2\hat{V}_{\textrm{wall}}}{i \omega} \frac{1}{\left[ \left(e^{k_f-k_s} - e^{-(k_f-k_s)}\right) \left( 1 - \alpha \right) - \left(e^{k_f + k_s} - e^{-(k_f + k_s)}\right) \left( 1 + \alpha \right)\right]}\\
    D &= -C
    % \lambda_2 \exp \left( -\lambda_2 L_s \right) \left[ 2c_1 + i \omega \mu_s \right]
    \end{aligned}
\end{equation}
We note here that \( k_f \) and \( k_s \) denote the fluid and solid wave contributions and \( \alpha \) represents the degree of fluid--solid coupling.
\section{Piecewise linear functions as particular solutions}\label{app:piecewise}
Here, we motivate the natural choice of using piecewise linear functions in our series solution for the
fluid and solid domains~(\mainref{eqn:sine_series}), in addition to the Fourier sine
series. Indeed, if we consider the simpler Couette flow case, i.e.~only viscous fluid between
two parallel plates (at \(L = 0\) and \(L = L_{\textrm{wall}}\),
with the top wall moving with \(V_{\textrm{wall}}\)), the \emph{trivial} solution lying in the {nullspace} of
the governing continuity and momentum equations (\mainref{eqn:simplified}) is \(v(y) = V_{\textrm{wall}}y/L_{\textrm{wall}}\).
Formally, this trivial solution is a particular solution of the governing linear PDE, arising due to an
inhomogeneity caused by a non-zero wall velocity (boundary conditions and
forcing terms are interchanged by Duhamel's principle). One can then view
\cref{eqn:sine_series} as a superposition of the particular solution and homogeneous
solution to the governing PDE.

\section{Useful identities in Fourier bases}\label{app:fourier_identities}
We proceed by utilizing the following orthogonality and integral relations for
Fourier basis functions, given \(k, l\neq 0\)
\begin{equation}
\label{eqn:fourier_relations}
\begin{aligned}
\textrm{Orthogonality relations}&
\quad
\begin{aligned}
    \int_{0}^{L} \sin\dfrac{\pi k y}{L} \sin\dfrac{\pi l y}{L} dy = \dfrac{L}{2}\delta_{kl} \\
    \int_{0}^{L} \cos\dfrac{\pi k y}{L} \cos\dfrac{\pi l y}{L} dy = \dfrac{L}{2}\delta_{kl}
\end{aligned}
\\
\textrm{Integral relations}&
\quad
\begin{aligned}
    \int_{0}^{L} \sin\dfrac{\pi k y}{L} = \dfrac{L}{\pi k} \left[ 1 - (-1)^k \right]  \\
    \int_{0}^{L} \dfrac{y}{L}\sin\dfrac{\pi k y}{L} = \dfrac{L}{\pi k}\left(- (-1)^k \right)
\end{aligned}
\end{aligned}
\end{equation}
where \(\delta_{kl}\) is the delta function, used in a pointwise sense.
\section{Details of neo-Hookean modal solution} \label{app:neo_hookean_modal_details}
Here we detail the closed-form modal solution in the case of a neo-Hookean solid.
We represent the solution \(u(y,t), v(y, t)\) using the Fourier sine series only in the upper half space \(y \geq 0\), as follows
\begin{equation}
    \label{eqn:app_sine_series}
    \begin{aligned}
        &v_{f}(\tilde{y}, t)=V_{I}(t)+\dfrac{\tilde{y}}{L_{f}}\left(V_{\textrm{wall}}(t)-V_{I}(t)\right)+\sum_{k=1}^{\infty} v_{f, k}(t) \sin \dfrac{\pi k \tilde{y}}{L_{f}} \\
        &u_{s}(y, t)=\dfrac{U_{I}(t) y}{L_{s}}+\sum_{k=1}^{\infty} u_{s, k}(t) \sin \dfrac{\pi k y}{L_{s}}
        % v_{s}(y, t)=\dfrac{V_{I}(t) y}{L_{s}}+\sum_{k=1}^{\infty} v_{s, k}(t) \sin \dfrac{\pi k y}{L_{s}},
    \end{aligned}
\end{equation}
where \(U_{I}(t), V_{I}(t)\) is the displacement and velocity of the solid--fluid interface at \(y = L_s\), \(\tilde{y} = y - L_s\), and \(v_{f,k}(t)\) and \(u_{s,k}(t)\) are the Fourier expansion coefficients of \( v_f \) and \(u_s\), respectively.
\begin{equation*}
    % \label{eq:temporal_form}
    \begin{aligned}
        V_{I}(t)    &= \imag{\hat{V}_{I} \exp (i \omega t)}\\
        v_{f, k}(t) &= \imag{\hat{v}_{f, k} \exp (i \omega t) }\\
        u_{s, k}(t) &= \imag{\hat{u}_{s, k} \exp (i \omega t) },
    \end{aligned}
\end{equation*}
with the immediate implication that
\begin{equation*}
    % \label{eq:temporally_transformed}
    \begin{aligned}
        U_{I}(t)    &= \imag{\frac{\hat{V}_{I}}{i \omega} \exp (i \omega t)} \\
        v_{s, k}(t) = \frac{\mathrm{d} u_{s, k}}{\mathrm{d} t} &= \imag{i \omega \hat{u}_{s, k} \exp (i \omega t) } \\
        \frac{\mathrm{d}^{2} u_{s, k}}{\mathrm{d} t^{2}} &= \imag{-\omega^2 \hat{u}_{s, k}
        \exp (i \omega t) }.
    \end{aligned}
\end{equation*}
Substitution of the temporal transformed quantities above in the momentum
ODEs and considering the boundary, interface conditions give the following
expression for \( \hat{v}_{f, k}, \hat{u}_{s, k}, \hat{V}_{I}\).
\begin{equation*}
    % \label{eq:linear_}
    \begin{aligned}
        \hat{v}_{f, k}&=\frac{\left\{(-1)^{k} \hat{V}_{\mathrm{wall}}-\hat{V}_{I}\right\} \alpha_{k}}{\pi k} \\
        \hat{u}_{s, k}&=-\frac{i(-1)^{k} \hat{V}_{I} \beta_{k}}{\pi \omega k} \\
        \hat{V}_{I}&=\frac{\hat{V}_{\mathrm{wall}} \left(1+\sum_{k=1}^{K-1}(-1)^{k} \alpha_{k}\right)}{\left(1+\sum_{k=1}^{K-1} \alpha_{k}\right) + \left(\frac{L_f}{L_s}\right) \left[ \left(\frac{\nu_s}{\nu_f}\right) \left(\frac{\rho_s}{\rho_f}\right)  - i \left(\frac{2c_{1}}{\omega \mu_f}\right) \right] \left(1+\sum_{k=1}^{K-1} \beta_{k}\right)}
    \end{aligned}
\end{equation*}
where
\begin{equation*}
    % \label{eq:linear_coeff}
    \begin{aligned}
        \alpha_{k} &= \frac{2}{1 - i \pi^2 k^2\left({L_{f}}^{-1}\left({\nu_{f}/\omega}\right)^{0.5}\right)^2} \\
        \beta_{k}  &= \frac{2}{1 - \pi^{2} k^{2}
        \left[ \left({\left(\omega L_s\right)}^{-1}\left(2c_1 / \rho_s \right)^{0.5} \right)^2 + i \left( {L_s}^{-1}\left(\nu_s / \omega\right)^{0.5} \right)^2 \right]}.
    \end{aligned}
\end{equation*}
The expressions above can then be directly used in~\cref{eqn:app_sine_series} to analytically evaluate solid displacements, fluid velocities and solid velocities, respectively.
\section{Comparison of direct analytical and modal solutions}\label{app:comparison}
\begin{figure}
\centering
\includegraphics[width=0.6\textwidth]{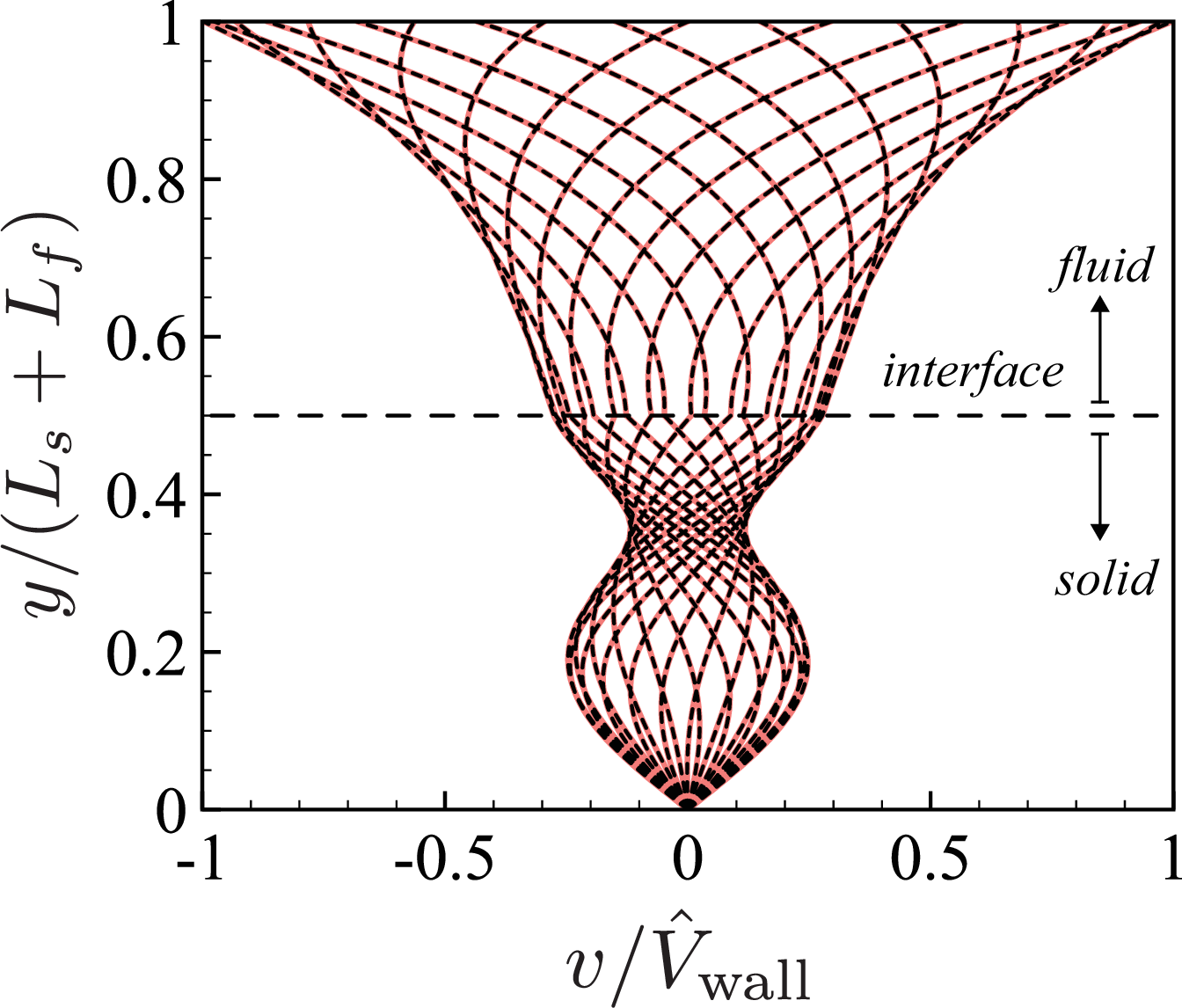}
\caption{
    \label{fig:comparison_solutions}
    Comparison of direct analytical and modal solutions. Plotting the non-dimensional velocity profiles for a visco-elastic solid with neo-Hookean model using the direct analytical (solid red line) and modal solutions (dashed black line) reveals good
    agreement across all times. The system is characterized by the same parameters as Figure 3 in the main text.
}
\end{figure}

Here, we showcase comparison between our direct analytical solutions discussed in~\cref{app:neo_hookean_details} and modal solutions discussed in~\cref{app:neo_hookean_modal_details} for a representative set of parameters. We overlay the velocity profiles obtained by these methods in ~\cref{fig:comparison_solutions} to see favorable comparison across all time instants. Hence our solutions, although obtained via different approaches, give the same results.
\section{Parameter details}\label{app:fig_details}
Here, for the purposes of repeatability and reproducibility, we document the parameter sets used to generate the results shown in the figures of our manuscript. We list the figure first and then tabulate the parameters used.

\subsection{Figure 2}
The parameters used in this figure are \(Re = 0.25 \), \(\Er = 1/\left(5\pi\right) \), \(\nu = 0\), \(\delta_f = 1.12\), \(\delta_s = 0\), \(\lambda = 2.52\), \( L = 2\), \(L_s = L_f = L / 4\), \(\rho_f = \rho_s = 1\), \(\mu_f = 1.0\), \(\mu_s = 0.0\), \(c_{1} = 2.5\), \(c_{3}=0\), \(\hat{V}_{\mathrm{wall}} = 1.0\), \(\omega = \pi\), \( \dot{\gamma} = \pi^{-1}\).

\subsection{Figure 3}
The system here is characterized by \(Re = 2, \Er = 1, \nu = 0.1, \delta_f = 0.4, \delta_s = 0.126, \lambda = 0.225\), \( L = 0.8\), \(L_s = L_f = L / 4 = 0.2 \), \(\rho_f = \rho_s = 1\), \(\mu_f = 0.02\), \(\mu_s = 0.1\mu_f\), \(c_{1} = 0.01\), \(c_{3}=0\), \(\hat{V}_{\mathrm{wall}} = 0.4\), \(\omega = \pi\), \( \dot{\gamma} = \pi^{-1}\).

The parameters used for the 2D DNS are \( \LCFL = 0.05, \CFL = 0.1\) with a resolution of \( [512]^2 \) in a domain of physical extent \( [0, 1]^2\). The reader is referred to~\citet{bhosale2021remeshed} for an interpretation of these parameters.

\subsection{Figure 6}
The system here is characterized by \(Re = 2, \Er = 1, \nu = 0.1, c = c_3/c_1 = 4, \delta_f = 0.4, \delta_s = 0.126, \lambda = 0.225\), \( L = 0.8\), \(L_s = L_f = L / 4 = 0.2 \), \(\rho_f = \rho_s = 1\), \(\mu_f = 0.02\), \(\mu_s = 0.1\mu_f\), \(c_{1} = 0.01\), \(c_{3}=0.04\), \(\hat{V}_{\mathrm{wall}} = 0.4\), \(\omega = \pi\), \( \dot{\gamma} = \pi^{-1}\).

The parameters used for the 2D DNS are \( \LCFL = 0.05, \CFL = 0.1\) with a resolution of \( [512]^2 \) in a domain of physical extent \( [0, 1]^2\). The reader is referred to~\citet{bhosale2021remeshed} for an interpretation of these parameters.
% \begin{table}
%   \begin{center}
% % \centering
%     \begin{tabular}{ll}
%     \toprule
%     Symbol & Value\\
%     \midrule
%     \( \Rey \) & \( 0.25 \) \\
%     \( \Er \) & \( 1/\left(5\pi\right) \) \\
%     \(\nu\) & \( 0 \) \\
%     \(\delta_f \) & \( 1.12 \) \\
%     \(\delta_s \) & \( 0 \) \\
%     \(\lambda \) & \( 2.52 \) \\
%     \( L \) & \( 2 \) \\
%     \( L_s \) & \( L / 4 = 0.5 \) \\
%     \( L_f \) & \( L / 4 = 0.5 \) \\
%     \( \rho_f \) & \( 1 \) \\
%     \( \rho_s \) & \( 1 \) \\
%     \(\mu_f \) & \( 1\) \\
%     \(\mu_s \) & \( 0\) \\
%     \(c_1 \) & \( 2.5\) \\
%     \(c_3 \) & \( 0\) \\
%     \(\hat{V}_{\mathrm{wall}} \) & \( 1\) \\
%     \(\omega \) & \(\pi\) \\
%     \( \dot{\gamma} \) & \(\pi^{-1}\) \\
%     \bottomrule
%     \end{tabular}
%     %\caption{Characteristic non-dimensional parameters}
%     % \label{tab:params}
%   \end{center}
% \end{table}
%
\section{Details on limit cases}\label{app:limit_cases}
%
% \subsection{No elastic solid~: single phase and multi-phase Stokes--Couette flows}
\begin{figure}
\centering
%height=.26\textheight
\includegraphics[width=\textwidth]{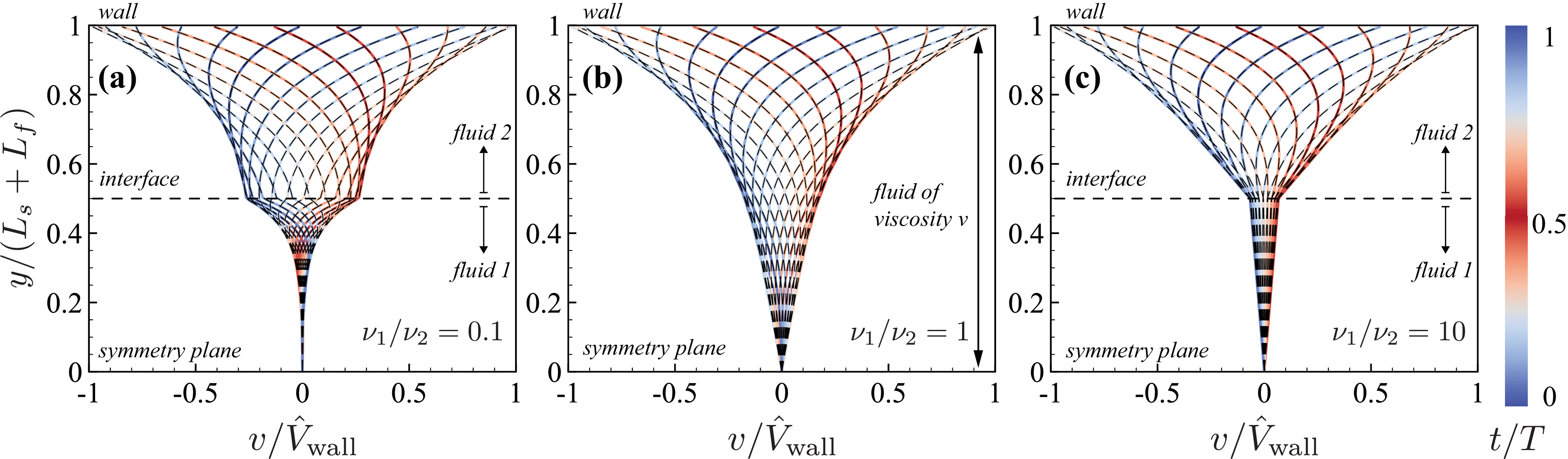}
\caption{\label{fig:noelastic}No elastic limit. Non-dimensional velocity profile in \(y\) when the solid is replaced by a viscous fluid. The system response is shown only for the upper half plane, for viscosity ratios \( \nu = \nu_1 / \nu_2 = \) \capsub{a} 0.1, \capsub{b} 1 and \capsub{c} 10. We compare our results with the reference Stokes--Couette solution~\cite{landau1987theoretical} in \capsub{b} and two-fluid Stokes--Couette solution~\cite{sim2006stratified,leclaire2014unsteady} in \capsub{a} and \capsub{c}. The reference solutions are shown in black dashed lines, and we find them in agreement with our solution. This system is characterized by \( Re  = 2, \delta = 0.4\). Other parameters used are \(L_s = L_f = L / 4 = 0.2\), \(\rho_f = \rho_s = 1\), \(\mu_f = 0.02\), \(c_{1} = 0\), \(c_{3}=0\), \(\hat{V}_{\mathrm{wall}} = 0.4\), \(\omega = \pi\). Colors represent \(t / T\).}
\end{figure}
Here, we present comparison with available analytical solutions in the limit of \(c_1 = 0\), as discussed in the main text. This limit indicates an absence of elastic forces in the solid phase. Hence, only viscous forces operate in the solid, effectively rendering it a
Newtonian fluid. First, if \( c_1 = 0 \) and \(\mu_s = \mu_f = \mu \) and \( \rho_s = \rho_f = \rho \),
i.e.~when the entire domain is occupied by a single fluid,
we recover the Stokes--Couette flow solution~\cite{landau1987theoretical} valid
throughout the domain. This is confirmed in~\cref{fig:noelastic}b, where we see
how our solution well agrees with the reference Stokes--Couette solutions,
depicted in black dashed lines. Details on the reference Stokes--Couette solution
can be found in~\cref{app:stokes_couette}.

Next, still within the \( c_1 = 0 \) limit but now with \(\mu_s \neq \mu_f\) or \( \rho_s \neq \rho_f\), i.e.~with the domain occupied by two different fluids, we
recover the multi-phase Stokes--Couette flow for two immiscible
liquids, which has established piecewise (in the two fluid domains) analytical
solutions~\cite{sim2006stratified,leclaire2014unsteady} found in~\cref{app:multiphase_stokes_coutte}. We showcase our
solutions for two viscosity ratios \( \nu = {\nu_1}/{\nu_2} = 0.1\) and
\( \nu = 10\) in~\cref{fig:noelastic}a,c, where subscripts \( 1, 2\) indicate
the bottom and top fluid respectively. Additionally, we overlay the reference solutions as
black dashed lines. As can be seen, we find favorable agreement across all times.

In all cases, \( \Rey = {\dot{\gamma} \omega L_2^2}/{\nu_2} = 2 \sim \order{1}\) and so we expect inertial and viscous contributions to be approximately equally important. Note that here the system Reynolds number is defined through the viscosity \( \nu_2 \) of fluid-2. Further, \(\delta_2 = 0.4 \) which indicates that boundary layer effects operate in the bulk of fluid-2. This leads to moderate velocity curvatures and magnitude decay in fluid-2 across~\cref{fig:noelastic}a,b,c. In the first case of~\cref{fig:noelastic}a, fluid-1 is ten times less viscous than fluid-2. This leads to relatively high velocity gradients of fluid-1 at the interface compared to fluid-2, reflecting the stress continuity condition~(\mainref{eqn:stress_continuity})
\begin{equation}
    \begin{aligned}
    \label{eq:gradjump}
        \frac{\partial_{y} v_{1}}{\partial_{y} v_{2}} = \frac{\nu_{2}}{\nu_{1}}.
    \end{aligned}
\end{equation}
The low viscosity of fluid-1 additionally implies a sharper fluid-1 boundary layer at the interface with \( \delta_1 = 0.13 \). Consequently, the velocity profiles have significant curvature within this layer in order for viscous forces to dissipate velocities to zero at the symmetry plane. In the second case~(\cref{fig:noelastic}b), where \(\nu_1 = \nu_2 \) implies a single fluid, the velocity profiles are smooth at all points in the domain and approach zero at the symmetry plane with a nearly linear profile outside the wall boundary layer (\(\delta = 0.4 \)). Meanwhile, in the third case of~\cref{fig:noelastic}c, fluid-1 is ten times more viscous than fluid-2. This time the gradient jump~(\cref{eq:gradjump}) across the interface in fluid-1 is comparatively smaller, which results in smaller fluid-1 oscillations. The high viscosity of fluid-1 leads to a boundary layer much larger than the fluid layer thickness, with \( \delta_1 = 1.26\). Hence, the velocity profiles are linear.
\subsection{Reference solution in the one fluid (no elastic solid) limit}\label{app:stokes_couette}
In the limit of \(c_1 = 0\) and \(\mu_s = \mu_f = \mu\) and \( \rho_s = \rho_f = \rho\),
i.e.~when the entire domain is occupied by a fluid of kinematic viscosity \(\nu\), we recover the
Stokes--Couette flow solution~\cite{landau1987theoretical} below
\[ v(y,t) = \hat{V}_{\mathrm{wall}} \imag{\frac{\sin(ky)}{\sin(k(L_s + L_f))}\exp (i \omega t)}; \quad k = \dfrac{1 - i}{\sqrt{2}} \sqrt{\frac{\omega}{\nu}}\]
valid throughout the domain. We showcased the comparison of this solution with our
results in~\cref{fig:noelastic}b.

\subsection{Reference solution in the two fluid (no elastic solid) limit}\label{app:multiphase_stokes_coutte}
In the limit of \(c_1 = 0\), but now with \(\mu_s \neq \mu_f\) or
\( \rho_s \neq \rho_f\), i.e. with the domain occupied by two different fluids, we
recover the multi-phase Stokes--Couette flow between two immiscible
liquids, which has established piecewise (in the two fluid domains) analytical solutions~\cite{sim2006stratified,leclaire2014unsteady}.

The solutions, for a setup with \( L_{1} = L_{2} = L / 2\), is given by
\begin{equation}
\begin{aligned}
v_{1}(\hat{y}, t) &= \hat{V}_{\mathrm{wall}} \imag{ \left( C_{12} \exp{\left( \lambda_{1}\hat{y} \right)} + C_{11} \exp{\left( -\lambda_{1}\hat{y} \right)} \right) \exp (i \omega t)} \quad \hat{y} \in [-1, 0) \\
v_{2}(\hat{y}, t) &= \hat{V}_{\mathrm{wall}} \imag{ \left( C_{22} \exp{\left( \lambda_{2}\hat{y} \right)} + C_{21} \exp{\left( -\lambda_{2}\hat{y} \right)} \right) \exp (i \omega t)} \quad \hat{y} \in [0, 1],
\end{aligned}
\end{equation}
where \( \hat{y} = 2y / L - 1 \) and
\begin{equation}
\begin{aligned}
\lambda_{1} &= (1 + i) \sqrt{\Rey_{1} / 2} \\
\lambda_{2} &= (1 + i) \sqrt{\Rey_{2} / 2},
\end{aligned}
\end{equation}
with \( \Rey_{1} = \dfrac{\omega L^2}{4\nu_{1}}, \Rey_{2} = \dfrac{\omega L^2}{4\nu_{2}} \). We determine
the constants \(C_{11}, C_{12}, C_{21}, C_{22} \) using the boundary and interface conditions, which leads
to the following system of equations
\begin{equation}
\begin{aligned}
\left(\begin{array}{cccc}
\mathrm{e}^{\lambda_{2}} & \mathrm{e}^{-\lambda_{2}} & 0 & 0 \\
1 & 1 & -1 & -1 \\
\lambda_{2} & -\lambda_{2} & M_{33} & M_{34} \\
0 & 0 & \mathrm{e}^{-\lambda_{1}} & \mathrm{e}^{\lambda_{1}}
\end{array}\right) \cdot\left(\begin{array}{c}
C_{22} \\
C_{21} \\
C_{12} \\
C_{11}
\end{array}\right)=\left(\begin{array}{c}
1 \\
0 \\
0 \\
0
\end{array}\right),
\end{aligned}
\end{equation}
with
\begin{equation}
\begin{aligned}
M_{33} &=-\frac{\Rey_{2}}{\Rey_{1}} \frac{\rho_{1}}{\rho_{2}} \lambda_{1} \\
M_{34} &=+\frac{\Rey_{2}}{\Rey_{1}} \frac{\rho_{1}}{\rho_{2}} \lambda_{1}
\end{aligned}
\end{equation}
which we then solve for. We showcased the comparison of this solution with our
results in~\cref{fig:noelastic}a,c. We remark that this setup is especially suited as a benchmark
for two-fluid simulations since the interface does not deform
(due to symmetry) and so curvature effects (such as those encountered while modeling
surface tension effects) are identically absent. Hence only the terms
contributing to interfacial stress jump are tested.
\section{Details on the semi-analytical solution for a generalized Mooney--Rivlin solid}
\label{app:mr_derivation}
% In the case of a generalized Mooney--Rivlin solid, with characteristic non-linear \(c_{3} \neq 0\)
% response to deformations, the governing equations exhibits nonlinearity which necessitates a
% numerical solution.
Here, we expand on the steps in deriving the semi-analytical solution for a generalized
Mooney--Rivlin solid,. To make the content clear and presentable, we repeat
some information provided in the main text.

In the case of a generalized Mooney--Rivlin solid, characterized by \(c_{3} \neq 0\), the hyperelastic stress
is proportional to the cubic power of strain in~\mainref{eqn:stress} which signifies a higher order (w.r.t strain)
non-linear response to deformations. In this case a semi-analytical solution, to be evaluated numerically, can be derived.
% Then \cref{eqn:mom_fluid_modal},
% \cref{eqn:mom_solid_modal} and \cref{eqn:stress_continuity_modal} can be solved
% analytically for the expansion coefficients.
% The second case, with  exhibits nonlinearity and needs a numerical solution.
First, we calculate the modal expansion coefficients of the nonlinear stresses \(\sigma_{\mathrm{NL}, k}\) in the governing equation~\cref{eqn:mom_solid_modal} from the main text. For this calculation we employ a Fourier
pseudospectral collocation scheme~\cite{sugiyama2011full}. The following cosine orthogonality
relation is useful in this case
\begin{equation}
\label{eqn:nonlinear_relations}
\begin{aligned}
    \int_{0}^{L} \cos\dfrac{\pi k y}{L} \cos\dfrac{\pi l y}{L} dy = \dfrac{L}{2}\delta_{kl}\left( 1 + \delta_{k0} \right)
\end{aligned}
\end{equation}
%
% Recollecting the in \cref{eqn:fourier_relations} we note that for \(k, l = 0\),
% the RHS in the above relation equals \(L\) or equivalently \(\frac{L}{2}
%    \delta_{kl}\left( 1 + \delta_{k0} \right)\).
%
Then upon transforming \mainref{eqn:nonlinear_stress} into the Fourier cosine bases and using~\cref{eqn:nonlinear_relations}, we obtain the following expression for \(\sigma_{\mathrm{NL}, k}\)
\begin{equation}
\label{eqn:stress_coefficients}
\sigma_{\mathrm{NL}, k} \approx \frac{8 c_{3}}{K\left(1+\delta_{k 0}\right)} \sum_{j=0}^{K-1}\left\{\frac{U_{I}}{L_{s}}+\sum_{l=1}^{K-1} \frac{\pi l u_{s, l}}{L_{s}} \cos \frac{\pi l\left(j+\frac{1}{2}\right)}{K}\right\}^{3} \cos \frac{\pi k\left(j+\frac{1}{2}\right)}{K}
\end{equation}
which is approximate, with numerical errors incurred from truncation (discussed in~\mainref{sec:fourier_series}) and our
choice of collocated quadrature, which is associated with the spatial discretization and sampling of the
nonlinear term \(\left({\partial u_{s}}/{\partial y}\right)^{3}\) at a
finite set of points \(x_j = (j + \dfrac{1}{2}) \Delta x\), with \(\Delta x
   = {L_s}/{K}\).

Next, we employ a numerical time integration scheme to evolve the non-linear~\cref{eqn:mom_fluid_modal,eqn:mom_solid_modal} from the main text. We use a second order constant timestepper comprised of mixed Crank-Nicolson (implicit, for stability in the viscous updates)
and explicit Nyström (midpoint rule) for the higher order time derivatives~\cite{hairer1991solving}.
The \(n^{\textrm{th}}\) time level at \(t = n \Delta t\) is denoted by a
superscript (\(n\)). First, the prescribed wall velocity is
\begin{equation}
    V_{\textrm{wall}}^{(n+1)} := V_{\textrm{wall}}((n+1)(\Delta t)) =
    \imag{\hat{V}_{\textrm{wall}} \exp (i \omega \left( (n+1)\Delta t \right) )}
\end{equation}
For the \(U_{I}\) update, which proceeds independently of the governing~\cref{eqn:mom_fluid_modal,eqn:mom_solid_modal} from the main text, we use the Crank-Nicolson scheme~\cite{hairer1991solving}, shown below
\begin{equation}
    \label{eq:displacement_update}
    U_{I}^{(n+1)} \approx U_{I}^{(n)}+\frac{\Delta t}{2}\left(V_{I}^{(n+1)}+V_{I}^{(n)}\right) + \order{\Delta t^2}
\end{equation}
We then turn our attention to the modal fluid momentum equation~(\mainref{eqn:mom_fluid_modal}).
For updating the fluid velocity modes with the action of the viscous terms,
we once again utilize the Crank-Nicolson discretization at time level \( n \)
\begin{equation}
    \begin{aligned}
        &\frac{2}{\pi k}\left\{ \frac{V_{I}^{(n + 1)} - V_{I}^{(n)}}{\Delta t}  -(-1)^{k} \frac{V_{\mathrm{wall}}^{(n+1)} - V_{\mathrm{wall}}^{(n)} }{\Delta t}\right\} +
        \frac{v_{f, k}^{(n + 1)} - v_{f, k}^{(n)} }{\Delta t} \\
        &= -\nu_{f}\left(\frac{\pi k}{L_{f}}\right)^{2} \frac{1}{2} \left( v_{f, k}^{(n+1)} + v_{f, k}^{(n)}\right) + \order{\Delta t^2}
    \end{aligned}
\end{equation}
where we note that both wall and interface velocities on the LHS are discretized consistently with \(v_{f, k} \).
Rearranging the terms in the equation above results in the following fluid mode update,
accurate up to second order in \( \Delta t \)
\begin{equation}
\label{eq:vsk}
    v_{f, k}^{(n+1)}=\frac{E_{f, k}}{\zeta_{f,k}}
\end{equation}
where
\begin{equation}
    % E_{f, k}=\left( 1 - \nu_{f} \dfrac{\Delta t}{2} \left(\dfrac{\pi k}{L_{f}}\right)^2 \right) v_{f, k}^{(n)} - \frac{2\left\{V_{I}^{(n+1)}-V_{I}^{(n)}-(-1)^{k} \delta V_{\mathrm{wall}}\right \}}{\pi k}
    E_{f, k}=\left( 2 - {\zeta_{f,k}} \right) v_{f, k}^{(n)} - \frac{2\left\{V_{I}^{(n+1)}-V_{I}^{(n)}-(-1)^{k} \delta V_{\mathrm{wall}}\right \}}{\pi k}
\end{equation}
\begin{equation}
    \zeta_{f, k} = 1 + \nu_{f} \dfrac{\Delta t}{2} \left(\dfrac{\pi k}{L_{f}}\right)^2
\end{equation}
with
\begin{equation}
    \delta V_{\textrm{wall}}=V_{\textrm{wall}}^{(n+1)}-V_{\textrm{wall}}^{(n)}
\end{equation}
Next, we focus on the modal solid momentum equation~(\mainref{eqn:mom_solid_modal}). Here, for updating solid displacements, we utilize the following explicit Nyström (midpoint rule) discretizations at the \(n^{\textrm{th}}\) time step
% \begin{equation}
% \label{eqn:mom_solid_modal}
% -\frac{2(-1)^{k}}{\pi k} \frac{\mathrm{d} V_{I}}{\mathrm{d} t}+\frac{\mathrm{d}^{2} u_{s, k}}{\mathrm{d} t^{2}} + \nu_s\left(\frac{\pi k}{L_{s}}\right)^{2} \frac{\mathrm{d} u_{s, k}}{\mathrm{d} t} + \frac{2c_{1}}{\rho_s}\left(\frac{\pi k}{L_{s}}\right)^{2} u_{s, k}+\frac{\pi k}{\rho_s L_{s}} \sigma_{\mathrm{NL}, k}=0
% \end{equation}
%
\[ \left( \frac{\mathrm{d} V_{I}}{\mathrm{d} t} \right)^{(n)} \approx \frac{ V_{I}^{(n+1)} - V_{I}^{(n-1)} }{2 \Delta t} + \order{\Delta t^2} \]
\[ \left( \frac{\mathrm{d} u_{s, k}}{\mathrm{d} t} \right)^{(n)} \approx \frac{ u_{s, k}^{(n+1)} - u_{s, k}^{(n-1)} }{2 \Delta t} + \order{\Delta t^2} \]
\[
\left( \frac{\mathrm{d}^{2} u_{s, k}}{\mathrm{d} t^{2}}\right)^{(n)}  \approx \frac{u_{s, k}^{(n+1)} - 2 u_{s, k}^{(n)} + u_{s, k}^{(n-1)}}{(\Delta t)^2} + \order{\Delta t^2}
\]
Upon substituting these discretizations in~\mainref{eqn:mom_solid_modal}, we arrive at
\begin{equation}
    \begin{aligned}
        &-\frac{2(-1)^{k}}{\pi k} \frac{ V_{I}^{(n+1)} - V_{I}^{(n-1)} }{2 \Delta t} + \frac{u_{s, k}^{(n+1)} - 2 u_{s, k}^{(n)} + u_{s, k}^{(n-1)}}{(\Delta t)^2} + \\
        & \nu_s\left(\frac{\pi k}{L_{s}}\right)^{2} \frac{u_{s, k}^{(n+1)} - u_{s, k}^{(n-1)}}{2 \Delta t} + \frac{2c_{1}}{\rho_s}\left(\frac{\pi k}{L_{s}}\right)^{2} u_{s, k}^{(n)} + \frac{\pi k}{\rho_s L_{s}} \sigma_{\mathrm{NL}, k}^{(n)} + \order{\Delta t^2} = 0
    \end{aligned}
\end{equation}
which upon algebraic manipulation leads to a second order temporally accurate solid displacement mode update
\begin{equation}
    \label{eq:usk}
    \begin{aligned}
        u_{s, k}^{(n+1)}=&\frac{1}{\zeta_{s,k}} \left[ \frac{(-1)^{k}\Delta t\left(V_{I}^{(n+1)}-V_{I}^{(n-1)}\right)}{\pi k} + 2 u_{s, k}^{(n)} - \left(2 - \zeta_{s,k}\right) u_{s, k}^{(n-1)} - \right. \\
        &\left.
        (\Delta t)^{2} \left(\frac{2c_{1}}{\rho_{s}} \left(\frac{\pi k}{L_{s}}\right)^2 u_{s, k}^{(n)}+\frac{\pi k}{\rho_{s} L_{s}} \sigma_{\mathrm{NL}, k}^{(n)}\right) \right]
    \end{aligned}
\end{equation}
with
\begin{equation}
    \zeta_{s, k} = 1 + \nu_{s} \dfrac{\Delta t}{2} \left(\dfrac{\pi k}{L_{s}}\right)^2
\end{equation}
Finally, we get a closed form for all the expressions by invoking the modal stress balance
~(\mainref{eqn:stress_continuity_modal}) at the \((n + 1)^{\mathrm{th}}\) time level below.
Our choice of discretization at the \((n + 1)^{\mathrm{th}}\) step, rather than at the
\(n^{\mathrm{th}}\) step, ensures that the updates to the solid~\cref{eq:usk} and
fluid~\cref{eq:vsk} modes are consistent with each other.
\begin{equation}
\label{eq:stress_balance_split}
\begin{aligned}
    &\left(\underbrace{\frac{\mu_{f}\left(V_{\textrm{wall}}-V_{I}\right)}{L_{f}} - \frac{2c_{1}U_{I}}{L_{s}} - \sigma_{\mathrm{NL}, 0} - \frac{\mu_{s}V_{I}}{L_{s}} }_{\text{\large \RN{1}}}\right)^{(n+1)} + \\
    &\left(\underbrace{\sum_{k=1}^{\infty}\left[\frac{\mu_{f} \pi k v_{f, k}}{L_{f}}-(-1)^{k}\left\{\frac{2c_{1} \pi k u_{s, k}}{L_{s}} +  \frac{\mu_{s}\pi k}{L_s}\frac{\mathrm{d} u_{s,k}}{\mathrm{d} t}+ \sigma_{\mathrm{NL}, k} \right\}\right]}_{\text{\large \RN{2}}}\right)^{(n+1)} = 0
\end{aligned}
\end{equation}
In the above~\cref{eq:stress_balance_split}, we first describe the time discretization of the first term, denoted by the under-brace \(\RN{1}\). For this term, the nonlinear contribution \(\sigma_{\mathrm{NL}}\) needs to be extrapolated
forward in time to the \((n + 1)^{\mathrm{th}}\) step. To achieve this, we utilize the second-order accurate extrapolation EXT2~\cite{hairer1991solving} scheme whose formula is shown below.
\begin{equation}
\label{eq:EXT2}
\sigma_{\mathrm{NL}, 0}^{(n+1)} \approx 2 \sigma_{\mathrm{NL}, 0}^{(n)} - \sigma_{\mathrm{NL}, 0}^{(n-1) } + \order{\Delta t^2}
\end{equation}
This discretization is physically motivated by the fact that the nonlinear terms govern the evolution of \emph{slow} non-linear elastic wave time scales \( \sim \order{{L_s}^{-1} \left(c_3 / \rho_s\right)^{0.5}} \) as opposed to the \emph{fast} diffusive time scales \(\sim  \order{L_s^{2}/\nu_s}\)  in the system. Hence, we can extrapolate these stress waves forward in time and still capture the correct physics. Upon substituting the discretizations of~\cref{eq:displacement_update,eq:EXT2} in the term \(\RN{1}\) discussed above, we arrive at
\begin{equation}
\label{eq:disc1}
\begin{aligned}
    \RN{1} \approx & \frac{\mu_{f}\left(V_{\textrm{wall}}^{(n+1)} - V_{I}^{(n+1)} \right)}{L_{f}} - \frac{2c_{1}}{L_{s}} \left( U_{I}^{(n)} + \frac{\Delta t}{2} \left( V_{I}^{(n+1)} + V_{I}^{(n)} \right) \right) - \\
    &
    (2\sigma_{\mathrm{NL}, 0}^{(n)} - \sigma_{\mathrm{NL}, 0}^{(n - 1)}) - \frac{\mu_{s}V_{I}^{(n + 1)}}{L_{s}}  + \order{\Delta t^2}
\end{aligned}
\end{equation}
% for some field quantity
% \(\dot{g}\). In our case, this field quantity \( g \) corresponds to the nonlinear
% term, i.e. \( g \equiv \sigma_{\mathrm{NL}} \).
%
Next, we discuss the time-discretization of the second term, denoted by the under-brace \(\RN{2}\) in~\cref{eq:stress_balance_split}. The following formulae are used for discretizing the terms in this case
\[  \left( \frac{\mathrm{d} u_{s,k}}{\mathrm{d} t} \right)^{(n+1)} \approx \frac{3u_{s,
    k}^{(n+1)} - 4u_{s, k}^{(n)} + u_{s, k}^{(n-1)}}{2 \Delta t} + \order{\Delta t^2} \]
\[  \sigma_{\mathrm{NL}, k}^{(n+1)} \approx 2 \sigma_{\mathrm{NL}, k}^{(n)} - \sigma_{\mathrm{NL}, k}^{(n-1) } + \order{\Delta t^2} \]
where the first discretization stems from the family of backward-difference formulae (BDF)~\cite{hairer1991solving} and the second discretization is similar to \cref{eq:EXT2}. Upon substituting these discretizations, the second term \(\RN{2}\) now reads
\begin{equation}
\label{eq:disc2}
\begin{aligned}
    \RN{2} \approx
    &\sum_{k=1}^{\infty}\left[
    \frac{\mu_{f} \pi k v_{f, k}^{(n+1)}}{L_{f}} \right. \\
    &\left.
    -(-1)^{k}
    \left\{
    \frac{2c_{1} \pi k u_{s, k}^{(n+1)}}{L_{s}} +  \frac{\mu_{s}\pi k}{L_s} \left( \frac{3u_{s,
    k}^{(n+1)} - 4u_{s, k}^{(n)} + u_{s, k}^{(n-1)}}{2 \Delta t}  \right)
    + 2 \sigma_{\mathrm{NL}, k}^{(n)} - \sigma_{\mathrm{NL}, k}^{(n-1) }
    \right\}
    \right] \\
    &+ \order{\Delta t^2}
\end{aligned}
\end{equation}
Now the terms at the \((n + 1)^{\mathrm{th}}\) step involving \(v_{f,k}, u_{s,k}\) can be directly substituted with the previously derived modal update expressions~(\cref{eq:usk,eq:vsk}). Following~\cref{eq:stress_balance_split} we set \( \RN{1} + \RN{2} = 0\), using their discretized versions in~\cref{eq:disc1,eq:disc2}. After standard (but tedious) algebraic manipulations, we finally arrive at the update equation for the interface velocity \( V_{I}^{(n+1)} \)
\begin{equation}
    V_{I}^{(n+1)} = \dfrac{E_{I}}{\dfrac{\mu_f}{L_f} \left( 1 + 2 \sum_{k=1}^{K-1} \dfrac{1}{\zeta_{f, k}}\right) + \dfrac{\mu_s}{L_s} \left( 1 + \dfrac{3}{2} \sum_{k=1}^{K-1} \dfrac{1}{\zeta_{s, k}}\right) + \dfrac{c_{1}\Delta t}{L_{s}}\left( 1 + 2 \sum_{k=1}^{K-1} \dfrac{1}{\zeta_{s, k}}\right) }
\end{equation}
where
% \begin{equation}
% \begin{aligned}
%     E_{I}=& \frac{\mu_{f} V_{\textrm{wall}}^{(n+1)}}{L_{f}} -  \frac{c_{1}}{L_{s}} \left\{ 2 U_{I}^{(n)}+(\Delta t) V_{I}^{(n)} \right\}  - 2 \sigma_{\mathrm{NL}, 0}^{(n)}+\sigma_{\mathrm{NL}, 0}^{(n-1)} \\
%     &+\sum_{k=1}^{K-1} \frac{\mu_{f}}{\zeta_{f,k} L_{f}} \left[ \pi k (2 - \zeta_{f, k}) v_{f, k}^{n} + 2 \left( V_{I}^{(n)}+(-1)^{k} \delta V_{\textrm{wall}} \right)  \right] \\
%     &+\sum_{k=1}^{K-1} \frac{2c_{1}}{\zeta_{s,k} L_{s}} \left[ V_{I}^{(n-1)}\Delta t + (-1)^{k} \pi k\left(\gamma_{k} u_{s, k}^{(n)} + (2 - \zeta_{s,k}) u_{s, k}^{(n-1)} \right) \right] \\
%     &+\sum_{k=1}^{K-1} \frac{\mu_{s}}{\zeta_{s,k} L_{s} \Delta t} \left[ \frac{3}{2} V_{I}^{(n-1)} \Delta t + (-1)^{k} \pi k \left( \left\{ \frac{3}{2}\gamma_{k} + 2 \zeta_{s, k}\right\} u_{s, k}^{(n)} + \left\{ \frac{3}{2}(2 - \zeta_{s,k}) - \frac{1}{2}\zeta_{s, k} \right\} u_{s, k}^{(n-1)} \right) \right] \\
%     &+\sum_{k=1}^{K-1}  (-1)^{k}\left( \left\{ \frac{\gamma_{k} + 2}{\zeta_{s, k}} + \frac{3}{2} \frac{\nu_{s} (\Delta t)}{\zeta_{s, k}} \left(\frac{\pi k}{L_{s}}\right)^2  - 2\right\} \sigma_{\mathrm{NL}, k}^{(n)} + \sigma_{\mathrm{NL}, k}^{(n-1)}\right)
%     % \left.+\left\{\right\}+\right]
% \end{aligned}
% \end{equation}
%
\begin{equation}
\begin{aligned}
    E_{I}=& \frac{\mu_{f} V_{\textrm{wall}}^{(n+1)}}{L_{f}} -  \frac{c_{1}}{L_{s}} \left\{ 2 U_{I}^{(n)}+(\Delta t) V_{I}^{(n)} \right\}  - 2 \sigma_{\mathrm{NL}, 0}^{(n)}+\sigma_{\mathrm{NL}, 0}^{(n-1)} \\
    &+\sum_{k=1}^{K-1} \frac{\mu_{f}}{\zeta_{f,k} L_{f}} \left[ \pi k (2 - \zeta_{f, k}) v_{f, k}^{n} + 2 \left( V_{I}^{(n)}+(-1)^{k} \delta V_{\textrm{wall}} \right)  \right] \\
    &+\sum_{k=1}^{K-1} \frac{2c_{1}}{\zeta_{s,k} L_{s}} \left[ V_{I}^{(n-1)}\Delta t + (-1)^{k} \pi k\left(\gamma_{k} u_{s, k}^{(n)} + (2 - \zeta_{s,k}) u_{s, k}^{(n-1)} \right) \right] \\
    &+\sum_{k=1}^{K-1} \frac{\mu_{s}}{\zeta_{s,k} L_{s} \Delta t} \left[ \frac{3}{2} V_{I}^{(n-1)} \Delta t + \right. \\
    &\left.
     (-1)^{k} \pi k \left( \left\{ \frac{3}{2}\gamma_{k} + 2 \zeta_{s, k}\right\} u_{s, k}^{(n)} +
    \left\{ \frac{3}{2}(2 - \zeta_{s,k}) - \frac{1}{2}\zeta_{s, k} \right\} u_{s, k}^{(n-1)} \right) \right] \\
    &+\sum_{k=1}^{K-1}  (-1)^{k}\left( \left\{ \frac{\gamma_{k} + 2}{\zeta_{s, k}} + \frac{3}{2} \frac{\nu_{s} (\Delta t)}{\zeta_{s, k}} \left(\frac{\pi k}{L_{s}}\right)^2  - 2\right\} \sigma_{\mathrm{NL}, k}^{(n)} + \sigma_{\mathrm{NL}, k}^{(n-1)}\right)
    % \left.+\left\{\right\}+\right]
\end{aligned}
\end{equation}
\begin{equation}
    \gamma_{k} = \frac{2c_{1}}{\rho_s}\left(\frac{\pi k \Delta t}{L_{s}}\right)^{2} - 2
\end{equation}
Overall, the above set of equations express the physical interface velocity, modal fluid velocities, modal solid displacements and physical interface displacement at a given time. Recovery of physical solid displacements and fluid velocities from their modal counterparts is achieved via~\cref{eqn:app_sine_series}. The system evolution can then be directly obtained by numerically iterating the above equations till the desired time. For convenience, the complete solution is given below.
\begin{equation*}
    \begin{aligned}
    % V_{I}^{(n+1)} = \frac{E_{I}}{\dfrac{\mu_f}{L_f} \left( 1 + 2 \sum_{k=1}^{K-1} \dfrac{1}{\zeta_{f, k}}\right) + \dfrac{\mu_s}{L_s} \left( 1 + \dfrac{3}{2} \sum_{k=1}^{K-1} \dfrac{1}{\zeta_{s, k}}\right) + \dfrac{c_{1}\Delta t}{L_{s}}\left( 1 + 2 \sum_{k=1}^{K-1} \dfrac{1}{\zeta_{s, k}}\right) }
    V_{I}^{(n+1)} &= {E_{I}} \biggr/ \left\{\dfrac{\mu_f}{L_f} \left( 1 + 2 \sum_{k=1}^{K-1} \dfrac{1}{\zeta_{f, k}}\right) + \dfrac{\mu_s}{L_s} \left( 1 + \dfrac{3}{2} \sum_{k=1}^{K-1} \dfrac{1}{\zeta_{s, k}}\right)
    % + \right. \\ &\left.
    + \dfrac{c_{1}\Delta t}{L_{s}}\left( 1 + 2 \sum_{k=1}^{K-1} \dfrac{1}{\zeta_{s, k}}\right)\right\} \\
% \end{equation*}
% \begin{equation*}
%     \end{aligned}
% \end{equation*}
% \begin{equation*}
%     \begin{aligned}
    u_{s, k}^{(n+1)}&=\frac{1}{\zeta_{s,k}} \left[ \frac{(-1)^{k}\Delta t\left(V_{I}^{(n+1)}-V_{I}^{(n-1)}\right)}{\pi k} + 2 u_{s, k}^{(n)} - \left(2 - \zeta_{s,k}\right) u_{s, k}^{(n-1)}
    - \right. \\
    &\left. (\Delta t)^{2} \left(\frac{2c_{1}}{\rho_{s}} \left(\frac{\pi k}{L_{s}}\right)^2 u_{s, k}^{(n)}+\frac{\pi k}{\rho_{s} L_{s}} \sigma_{\mathrm{NL}, k}^{(n)}\right) \right] \\
    \end{aligned}
\end{equation*}
\begin{equation*}
    \begin{aligned}
        v_{f, k}^{(n+1)}=\frac{E_{f, k}}{\zeta_{f,k}} ; \qquad
        U_{I}^{(n+1)}=U_{I}^{(n)}+\frac{\Delta t}{2}\left(V_{I}^{(n+1)}+V_{I}^{(n)}\right) ;
    \end{aligned}
\end{equation*}
with
\begin{equation*}
    \begin{aligned}
    \zeta_{s, k} = 1 + \nu_{s} \dfrac{\Delta t}{2} \left(\dfrac{\pi k}{L_{s}}\right)^2 ;
    \qquad
    \zeta_{f, k} = 1 + \nu_{f} \dfrac{\Delta t}{2} \left(\dfrac{\pi k}{L_{f}}\right)^2 ; \\
    \delta V_{\textrm{wall}} = V_{\textrm{wall}}^{(n+1)}-V_{\textrm{wall}}^{(n)} ;
    \qquad
    \gamma_{k} = \frac{2c_{1}}{\rho_s}\left(\frac{\pi k \Delta t}{L_{s}}\right)^{2} - 2 ; \\
    % E_{f, k}=\left( 1 - \nu_{f} \dfrac{\Delta t}{2} \left(\dfrac{\pi k}{L_{f}}\right)^2 \right) v_{f, k}^{(n)} - \frac{2\left\{V_{I}^{(n+1)}-V_{I}^{(n)}-(-1)^{k} \delta V_{\mathrm{wall}}\right \}}{\pi k}
    \end{aligned}
\end{equation*}
\begin{equation*}
    \begin{aligned}
        E_{f, k} &=\left( 2 - {\zeta_{f,k}} \right) v_{f, k}^{(n)} - \frac{2\left\{V_{I}^{(n+1)}-V_{I}^{(n)}-(-1)^{k} \delta V_{\mathrm{wall}}\right \}}{\pi k} \\
        E_{I}=& \frac{\mu_{f} V_{\textrm{wall}}^{(n+1)}}{L_{f}} -  \frac{c_{1}}{L_{s}} \left\{ 2 U_{I}^{(n)}+(\Delta t) V_{I}^{(n)} \right\}  - 2 \sigma_{\mathrm{NL}, 0}^{(n)}+\sigma_{\mathrm{NL}, 0}^{(n-1)} \\
        &+\sum_{k=1}^{K-1} \frac{\mu_{f}}{\zeta_{f,k} L_{f}} \left[ \pi k (2 - \zeta_{f, k}) v_{f, k}^{n} + 2 \left( V_{I}^{(n)}+(-1)^{k} \delta V_{\textrm{wall}} \right)  \right] \\
        &+\sum_{k=1}^{K-1} \frac{2c_{1}}{\zeta_{s,k} L_{s}} \left[ V_{I}^{(n-1)}\Delta t + (-1)^{k} \pi k\left(\gamma_{k} u_{s, k}^{(n)} + (2 - \zeta_{s,k}) u_{s, k}^{(n-1)} \right) \right] \\
        &+\sum_{k=1}^{K-1} \frac{\mu_{s}}{\zeta_{s,k} L_{s} \Delta t} \left[ \frac{3}{2} V_{I}^{(n-1)} \Delta t + \right. \\
        &\left.
         (-1)^{k} \pi k \left( \left\{ \frac{3}{2}\gamma_{k} + 2 \zeta_{s, k}\right\} u_{s, k}^{(n)} +
        \left\{ \frac{3}{2}(2 - \zeta_{s,k}) - \frac{1}{2}\zeta_{s, k} \right\} u_{s, k}^{(n-1)} \right) \right] \\
        &+\sum_{k=1}^{K-1}  (-1)^{k}\left( \left\{ \frac{\gamma_{k} + 2}{\zeta_{s, k}} + \frac{3}{2} \frac{\nu_{s} (\Delta t)}{\zeta_{s, k}} \left(\frac{\pi k}{L_{s}}\right)^2  - 2\right\} \sigma_{\mathrm{NL}, k}^{(n)} + \sigma_{\mathrm{NL}, k}^{(n-1)}\right).
        % \left.+\left\{\right\}+\right]
    \end{aligned}
\end{equation*}
\section{Online sandbox for interactive simulations}\label{app:sandbox}
The results presented in~\cref{sec:nh_behavior,sec:mr_behavior} in the main text serve as a minimal platform
for exploring flow--elastic structure interaction phenomena, thus building intuition
into more complex problems. To aid this exploration, we
open-source our computational code under a liberal license. Further, to enable a seamless
research/educational experience and to disencumber scientists/students
from the process of installing essential computational software stack, we provide
an interactive sandbox built atop our code. This sandbox is free, open-source, hosted
online and is accessible from any modern web browser running on personal devices
from mobile phones to laptops. Our sandbox can be found \href{https://gazzolalab.github.io/parallel_slab_sandbox/parallel_slab_sandbox.html}{here} and the numerical code powering it is available \href{?}{here}.
% Snapshots of the sandbox can be % found in~\cref{app:sandbox}.
% \section{Online sandbox for interactive simulations}\label{app:sandbox}
We present snapshots of this sandbox in~\cref{fig:online_sandbox}. Users can
utilize simple sliders to change geometrical and dynamical parameters
on the left, which are then used to run computations asynchronously before
presenting results and velocity plots on the right.
\begin{figure}
\centering
\includegraphics[width=\textwidth]{images/online_sandbox.eps}
\caption{
    \label{fig:online_sandbox}
    Online sandbox. Using sliders to control the geometrical and dynamical parameters, users can investigate
    the response of the system for a wide range of parameters, shown for two representative cases. Here the settings of \capsub{a} corresponds to the \( \Rey = 2, \Er = 1\) case, whose profiles were discussed earlier in~\cref{fig:neo_hookean_simulations}. Changing the solid properties (elasticity via \( \Er\), occupancy via \( L_f / L_s \) and density via \( \rho \)) leads to the profiles shown in \capsub{b}. In both cases, the interface is marked with a solid, black line for visual clarity.
}
\end{figure}
%
%\bibliographystyle{jfm}
%\bibliography{parallel_slab_lit.bib}